\documentclass[
  aps,
  prd,
  preprint,
  preprintnumbers,
  nofootinbib,
  nobibnotes, longbibliography,
  superscriptaddress
]{revtex4-1}

\usepackage{amssymb,amsmath,bm,natbib}
\usepackage[dvipsnames]{xcolor}
\usepackage[colorlinks = true,
linkcolor = Maroon,
urlcolor  = Maroon,
citecolor = Maroon]{hyperref}
\usepackage{microtype}
\usepackage{graphicx}
\usepackage{xspace}
\usepackage[normalem]{ulem}
\usepackage{dsfont}
\usepackage{slashed}
\usepackage{caption}
\usepackage{subcaption}
\usepackage{comment}

\newcommand{\ygn}[1]{{\bf \color{red} [YG:~#1]}}

\newcommand{\Kbar}{{\overline{K}^0}}

\newcommand\bra[2][]{#1\langle {#2} #1\rvert}
\newcommand\ket[2][]{#1\lvert {#2} #1\rangle}

\newcommand{\mycomment}[1]{}

\newcommand{\Cornelladdr}{Department of Physics, LEPP, Cornell University, Ithaca, NY 14853, USA}

\newcommand{\DESYddr}{Deutsches Elektronen–Synchrotron, 22607 Hamburg, Germany}
\begin{document}

\title{$CP$ violation in decays into states with neutral kaons
}

\author{Yuval~Grossman}
\email{yg73@cornell.edu}
\affiliation{\Cornelladdr}
%Department of Physics, LEPP, Cornell University, Ithaca, NY 14853, USA}

\author{Paolo~Leo}
\email{paolo.leo@desy.de}
\affiliation{\DESYddr}

\author{Alberto~Martini}
\email{alberto.martini@desy.de}
\affiliation{\DESYddr}

\author{Guglielmo~Papiri}
\email{gp343@cornell.edu}
\affiliation{\Cornelladdr}

\author{Armine~Rostomyan}
\email{armine.rostomyan@desy.de}
\affiliation{\DESYddr}

\date{\today}

\begin{abstract}
$CP$  violation in the kaon system can manifest itself in decays to final states containing neutral kaons. The effect is governed by an efficiency function that reflects the specific experimental setup. We demonstrate that this efficiency function factorizes into two components: the kaon energy spectrum and a universal detector‑dependent factor. We show that matter effects on kaon oscillations can have a significant effect on $CP$  asymmetries. We provide estimates of the theoretical prediction for some such $CP$  asymmetries for a Belle~II–type experiment.

%We show that the measured time-integrated $CP$  asymmetry is influenced by the experimental detection efficiency as a function of both the energy in the laboratory frame and the decay time of the kaon\gp{This needs to be changed}. This implies a non-vanishing $CP$  asymmetry for measurements in the decay channel $\tau\rightarrow\pi K_S K_L \nu$. We derive a theoretical prediction for this asymmetry and discuss its experimental relevance.
\end{abstract}

\maketitle

%%%%%%%%%%%%%%%%%%%%%%%%%%%%%%%%%%%%%%%%%%%%%%%%%%%%%%%%%%%

\section{Introduction}

%\gp{Maybe an opening motivation like: The study of $CP$  violation in semileptonic $\tau$ decays is important to probe direct $CP$  violation in the lepton sector and investigate physics beyond the Standard Model.}
%The $CP$  \desy{asymmetry} in the $\tau\rightarrow \pi K_S\nu$ decay,
%\begin{equation}\label{A1_general}
%A_1 = \frac{ \Gamma(\tau^+\rightarrow \pi^+ K_S\overline{\nu_{\tau}}) - \Gamma(\tau^-\rightarrow \pi^- K_S\nu_{\tau})} {\Gamma(\tau^+\rightarrow \pi^+ K_S\overline{\nu_{\tau}}) + \Gamma(\tau^-\rightarrow \pi^- K_S\nu_{\tau})}\;,
%\end{equation}
%has been investigated by the BaBar~\cite{BaBar:2012} and Belle~\cite{BELLE:2011} collaborations. Within the Standard Model (SM) the origin of the $CP$  asymmetry is that $\tau^+$ and $\tau^-$ decay into the two different kaon interaction eigenstates $K^0$ and $\overline{K^0}$, which have different projection into the mass eigenstates $K_S$ and $K_L$.  

In the standard model (SM), the $CP$  asymmetry in the $\tau$ decay 
\begin{equation}\label{A1_general}
A_1 = \frac{ \Gamma(\tau^+\rightarrow \pi^+ K_S\overline{\nu_{\tau}}) - \Gamma(\tau^-\rightarrow \pi^- K_S\nu_{\tau})} {\Gamma(\tau^+\rightarrow \pi^+ K_S\overline{\nu_{\tau}}) + \Gamma(\tau^-\rightarrow \pi^- K_S\nu_{\tau})}\;,
\end{equation}
arises because $\tau^+$ and $\tau^-$  decay to different kaon interaction eigenstates ($K^0$ and ${\overline{K}^0}$), which project differently onto the mass eigenstates ($K_S$ and $K_L$). 
The first theoretical estimate for the $CP$ asymmetry $A_1$  was given in Ref.~\cite{Bigi:2005},
\begin{equation}\label{A1_Bigi_Sanda}
    A_1 \approx 2\text{Re}[\epsilon] \approx 3.3\times 10^{-3}\;,
\end{equation}
where $\epsilon$ is the $CP$  violation parameter measured in neutral kaon mixing. This result was derived assuming that the final state is a $K_S$ eigenstate. It is important to note that experimentally the $K_S$ in the final state is defined as the $\pi\pi$ system with invariant mass $m_{\pi\pi} \approx m_K$. %This differs slightly from the true KS eigenstate, and the mismatch, particularly due to KL-KS interference, affects the measured asymmetry.
Later, in~\cite{Grossman:2011}, this prediction was revisited showing that the fact that a $K_S$ is identified by its decay to two pions introduces important effects due to $K_L-K_S$ interference. As a result, $A_1$ depends on the decay time of the kaon and the experimental requirements. 

The $CP$ asymmetry $A_1$ has been investigated by the BaBar~\cite{BaBar:2012} and Belle~\cite{BELLE:2011} collaborations.  The BaBar decay-rate asymmetry result shows a $3\sigma$ tension with the SM prediction, while Belle’s angular analysis finds results consistent with zero within $O(10^{-2})$ per $K_S$ mass bin. 
%The measurements indicate a discrepancy as they are consistent with the available SM prediction only at the $3\sigma$ level. 

%It is important to note that the $K_S$ in the final state is defined as the $\pi\pi$ system with invariant mass $m_{\pi\pi} \approx m_K$. In what follows, we continue to use this definition; however, we emphasize that the deviation between the true $K_S$ state and the state that decays into two pions is crucial for understanding the effects discussed here.

The Belle~II collaboration~\cite{BELLE2:2018} aims to improve the measurement precision. 
However,  the observed asymmetry in $e^+e^- \to \tau^+\tau^-$ events includes background contributions from other $\tau$ decays, particularly those involving two kaons, $\tau^\pm \to K^\pm K_S \, \nu_\tau$ and  $\tau^\pm \to \pi^\pm K_S K_L \nu_\tau$, with corresponding asymmetries defined as:
\begin{equation}\label{A2}
A_2 = \frac{\Gamma(\tau^+\rightarrow \overline{\nu_{\tau}} K^+ K_S) - \Gamma(\tau^-\rightarrow \nu_{\tau} K^- K_S)}{\Gamma(\tau^+\rightarrow \overline{\nu_{\tau}} K^+ K_S) + \Gamma(\tau^-\rightarrow \nu_{\tau} K^- K_S)}\; , 
\end{equation}
and
\begin{equation}\label{A3}
A_3 = \frac{\Gamma(\tau^+\rightarrow \overline{\nu_{\tau}} K_L K_S \pi^+) - \Gamma(\tau^-\rightarrow \nu_{\tau} K_L K_S \pi^-)}{\Gamma(\tau^+\rightarrow \overline{\nu_{\tau}} K_L K_S \pi^+) + \Gamma(\tau^-\rightarrow \nu_{\tau} K_L K_S \pi^-)}\; .
\end{equation}
For an ideal experiment we expect $A_1 = -A_2$ and $A_3=0$. However, as pointed out in Ref.~\cite{Grossman:2011}, in general the experimentally measured $CP$  asymmetries depend on an efficiency function,  describing the probability to reconstruct a $K\rightarrow\pi\pi$ decay as a function of the kaon rest-frame decay time, $t$.  Thus, these equalities are violated in a real experiment. 
Ref.~\cite{Grossman:2011} treated the efficiency function as  universal across decay modes. However, it depends on the specific process from which the kaon originates.

%In Ref.~\cite{Grossman:2011} the dependence of the asymmetry $A_1$ was assumed to be only on the cuts in the kaon rest frame decay time. This assumption was also was used in the BaBar analysis~\cite{BaBar:2012}. In this work we show that the experimental cuts depend also on the energy distribution of the kaon in the laboratory frame, that is specific to the decay process considered. While the effect of the assumption was below the errors in previous experiments, that may not be the case in the future. 

%the iscussion were done about how to it was assumed that this function only depends on the feature of the specific experiment reconstruction analysis, and depends only on the kaon decay time in its rest frame. As we discuss below, this simplification is not justified in the general case. 

In this work, we extend that analysis by showing that 
%aim to extend the analysis presented in Ref.~\cite{Grossman:2011} regarding the rule of the efficiency function $f_P(t)$ and its application for realistic experiments. We %discuss the dependence of $f_P(t)$ on the energy, momentum, and decay time distributions of the kaons emitted in the specific process $P$, and we 
%show that $f_P(t)$ depends on the decay process. 
%This can lead to different values of the $CP$  asymmetry. 
%In particular, 
%show that 
the reconstruction efficiency function  factorizes into two fundamental parts: a universal, detector-specific efficiency in the laboratory frame,  and process-dependent kaon energy spectrum. Moreover, we show that matter effects on neutral kaon oscillations can give very important corrections to the measured $CP$  asymmetries.
%As we explain below, for an ideal experiment we expect $A_3=0$ and $A_1 = -A_2$. However, the fact that the efficiency function depends on the decay process, makes these equalities violated in a realistic experiment. We provide theoretical predictions for these asymmetries in the case of a Belle~II-type experiment and discuss the validity of such approximations. 
%we focus on the $\tau\rightarrow\pi K_S K_L\nu$ decay mode. Such decay process originates from the $\tau\rightarrow\pi K^0 \Kbar\nu$ decay where the kaon and anti-kaon ended up decaying as a $K_S$ and $K_L$.  In an ideal experiment, the time-integrated asymmetry due to $K^0$ and $\Kbar$ cancel each other between the $K_S$ and the $K_L$ decays. We show that the time and energy dependence of the reconstruction efficiency makes $A_3\neq 0$ in a realistic experiment. We provide an upper bound for the asymmetry $A_3$ on Belle~II. We then discuss the importance of such background in the experimental analysis to measure $A_1$.
We provide theoretical predictions for all these asymmetries in the case of a Belle~II-type experiment. 
We show that, given  the anticipated precision at Belle~II, it is safe to assume $A_3=0$ and $A_1=-A_2$, since  the associated uncertainties are negligible compared to other sources of uncertainty. In contrast, corrections arising from the matter effects are subject to significant theoretical uncertainties and therefore should be studied  within  the experimental analysis.

%%%%%%%%%%%%%%%%%%%%%%%%

%%%%%%%%%%%%%%%%%%
\section{The experimental asymmetry}
In general,  the experimentally measured $CP$  asymmetry $A_{Exp,P}$ associated with a given $\tau$ decay process $P$ receives contributions from three distinct sources:
\begin{enumerate} 
\item The intrinsic (direct) $CP$ asymmetry in the $\tau$ decay, denoted $A_{Exp,P}^{\tau}$;
\item The contribution from $CP$ violation in the kaon system, arising from neutral-kaon oscillations in the final state, denoted $A_{Exp,P}^{K}$;
\item The detector-induced asymmetry, $A_{Exp,P}^{\text{det}}$, originating from possible differences in the detection efficiencies of the final-state particles in $\tau^+$ versus $\tau^-$ decays.
\end{enumerate}
The first two contributions are genuine $CP$-violating effects, while the last term represents an artificial asymmetry induced by the detector.
Assuming that all contributions are small, they can be treated additively:
\begin{equation}\label{A1}
A_{Exp,P}=A_{Exp,P}^{\tau} +  A_{Exp,P}^{K} + A_{Exp,P}^{\rm det} \;.
\end{equation}
Within the SM, the first contribution, $A_{Exp,P}^{\tau}$, is negligible.  Accordingly, in the following discussion we set $A_{Exp,P}^{\tau}=0$.  This work focuses on the second and third contributions to the asymmetry, which are related to the experimental reconstruction efficiency for detecting the kaons. 

%The sample receives contributions from three decays:
%\begin{enumerate}
%    \item $\tau^\pm \to \pi^\pm K_S^0 (\geq 0\pi^0)\, \nu_\tau$,
%    \item $\tau^\pm \to K^\pm K_S^0 (\geq 0\pi^0)\, \nu_\tau$,
%    \item $\tau^\pm \to \pi^\pm K^0 \bar{K}^0 (\geq 0 \pi^0)\nu_\tau$. 
%\end{enumerate}
%The first two channels involve \KS and are sensitive to CP violation effects arising from the neutral kaon system sector. The third decay mode does not induce CP violation within the SM. However, it is an irreducible background, since it can mimic the signal topology.

\subsection{The kaon frame efficiency}

We define $f(t)$ to be the experimental reconstruction efficiency function for the detection of a neutral kaon in the final state of the $\tau$ decay. Here $t$ is the kaon decay time in the kaon rest frame. In general, $f(t)$ carries 3 indices: the process $P$ from which the kaon originated, $P=1,2,3$ (related to the asymmetries in Eqs. (\ref{A1_general}), (\ref{A2}) and (\ref{A3}));  the charge of the $\tau$; and the strangeness of the kaon, that is,  $K^0$ or $\overline{K}^0$. Thus, in general, for any process $P$ there are four $f_P(t)$ functions: 
\begin{equation}\label{f_P_ALL}
f_P^{+,K}, \qquad
f_P^{+,\overline{K}}, \qquad
f_P^{-,K}, \qquad
f_P^{-,\overline{K}}.
\end{equation}
If the process $P$ has only one neutral kaon in final state, as for $A_1$ and $A_2$, then two of the above functions are irrelevant. Indeed, the decay processes written in terms of kaons interaction eigenstates are:
\begin{align}
    A_1:&\quad \tau^+\rightarrow \overline{\nu}\pi^+K^0\;,\\
    A_2:&\quad \tau^+\rightarrow \overline{\nu}K^+\overline{K}^0\;,
\end{align}
meaning that we do not need to consider
$f_1^{-,K}$, $f_1^{+,\overline{K}}$, $f_2^{+,K}$, and $f_2^{-,\overline{K}}$.
For decays with two neutral kaons in final states, as in the case of $A_3$, all four combinations are relevant.

We further define
\begin{align}\label{f_sum}
f_P^K(t) = \frac{f_P^{+,K}(t)+f_P^{-,\overline{K}}(t)}{2},\qquad
\Delta f_P^K(t) = \frac{f^{+,K}_P(t)-f_P^{-,\overline{K}}(t)}{2}, \\
f_P^{\overline{K}}(t) = \frac{f_P^{+,\overline{K}}(t)+f_P^{-,K}(t)}{2},\qquad
\Delta f_P^{\overline{K}}(t) = \frac{f^{+,\overline{K}}_P(t)-f_P^{-,K}(t)}{2} ,
\end{align}
and the functions with no superindex
\begin{equation}\label{f_P_FULL_FULL}
    f_P(t)=f_P^K(t) + f_P^{\overline{K}}(t), \qquad\Delta f_P(t)=f_P^K(t) - f_P^{\overline{K}}(t)\;.
\end{equation}
For the case of $A_1$ we can make the kaon index implicit
\begin{equation}
    f_1^K(t) = f_1(t)\;,\qquad f_1^{\overline{K}}(t)=0\;,
\end{equation}
and similarly for $A_2$
\begin{equation}
    f_2^{\overline{K}}(t)=f_2(t)\;,\qquad f_2^K(t)=0\;,
\end{equation}
where we have set $f_1^{\overline{K}}$ and $f_2^K$ to zero, as they are irrelevant for the relative processes.  In the case of $A_3$, as we discuss in Section~\ref{sec:A3}, both $f_3^{K}$ and $f_3^{\overline{K}}$ are relevant. 

\subsection{$CP$  asymmetry for decays with one neutral kaon}

We start by considering the $CP$  asymmetry for a process $P$ with a single neutral kaon in the final state. For the sake of simplicity we develop the formalism for $A_1$. %, that is we consider $P=1$.

Under the assumptions stated above, the measured asymmetry is given by
\begin{align}\label{A_Exp}
A_{Exp,1}=\frac{\int_{0}^{\infty}dt\;[f_1^{+,K}(t)\Gamma_{\pi\pi}(t) - f_1^{-,\overline{K}}(t)\overline{\Gamma}_{\pi\pi}(t)]}{\int_{0}^{\infty}dt\;[f_1^{+,K}(t)\Gamma_{\pi\pi}(t) + f_1^{-,\overline{K}}(t)\overline{\Gamma}_{\pi\pi}(t)]}\;.
\end{align}
%where
%, $Exp$ labels the fact that we refer to the experimental measured asymmetry, 
%$f_P^K(t)$ ($f_{\overline{P}}^{\overline{K}}$) is the experimental reconstruction efficiency functions for the $K^0$ ($\Kbar$) detection, and $P$ ($\Pbar$) is a subscript that indicates the process from which the kaons originate.
The rates, 
\begin{align}\label{gamma_gammaBar}
\Gamma_{\pi\pi}(t) = \Gamma(K^0(t)\rightarrow2\pi), \qquad
\overline{\Gamma}_{\pi\pi}(t) = \Gamma(\overline{K}^0(t)\rightarrow 2\pi),
\end{align}
are defined as the decay rates for $K^0$ and $\Kbar$ into two pions, as a function of the decay time $t$ in the reference frame of the kaon. These rates  %(\ref{gamma_gammaBar}) 
are known and are expressed in terms of measured physical parameters, 
see Eq.~(\ref{Gamma_Full}) and Eq.~(\ref{GammaBar_Full}).
We define
\begin{equation}\label{Delta_Sigma_pi_pi}
\Sigma_{\pi\pi}(t)=\Gamma_{\pi\pi}(t) + \overline{\Gamma}_{\pi\pi}(t),\qquad \Delta_{\pi\pi}(t)=\Gamma_{\pi\pi}(t) - \overline{\Gamma}_{\pi\pi}(t)\;.
\end{equation}
We know experimentally that 
$\Delta_{\pi\pi}(t)/\Sigma_{\pi\pi}(t) = O(\epsilon)$. In addition, in all available experiments $\Delta f^K_1(t) \ll f_1^K(t)$. Thus we can treat both $\Delta_{\pi\pi}(t)$ and $\Delta f_1^K(t)$ as small.
Then, Eq.~(\ref{A_Exp}) can be written to leading order in the above small parameters  as
\begin{equation}
A_{Exp,1}\approx \frac{\int_0^{\infty}dt[\Delta f_1^K(t)\Sigma_{\pi\pi}(t) + f_1^K(t)\Delta_{\pi\pi}(t)]}{\int_0^{\infty}dt\;f_1(t)\Sigma_{\pi\pi}(t)}\equiv A_{Exp,1}^{\text{det}} + A_{Exp,1}^K\;,
\end{equation}
where we define:
\begin{equation}\label{A1_Final}
A_{Exp,1}^{\text{det}}=\frac{\int_0^{\infty}dt\;\Delta f_1^K(t)\Sigma_{\pi\pi}(t)}{\int_0^{\infty}dt\;f_1^K(t)\Sigma_{\pi\pi}(t)}\,,\qquad A_{Exp,1}^{K}=\frac{\int_0^{\infty}dt\; f_1^K(t)\Delta_{\pi\pi}(t)}{\int_0^{\infty}dt\;f_1^K(t)\Sigma_{\pi\pi}(t)}\;.
\end{equation}

The result for $A_2$ can be obtained from Eq.~(\ref{A_Exp}) by  $K \leftrightarrow \overline{K}$ and $P=2$. This leads to
 \begin{equation}\label{A2_Final}
A_{Exp,2}^{\text{det}}=-\frac{\int_0^{\infty}dt\;\Delta f_2^{\overline{K}}(t)\Sigma_{\pi\pi}(t)}{\int_0^{\infty}dt\;f_2^{\overline{K}}(t)\Sigma_{\pi\pi}(t)}\,,\qquad A_{Exp,2}^{K}=-\frac{\int_0^{\infty}dt\; f_2^{\overline{K}}(t)\Delta_{\pi\pi}(t)}{\int_0^{\infty}dt\;f_2^{\overline{K}}(t)\Sigma_{\pi\pi}(t)}\;.
\end{equation}
The situation for $A_3$ is more involved and it is discussed in Section~\ref{sec:A3}.

As defined before, $A_{Exp,P}^{\text{det}}$ is the asymmetry due to the different detection efficiency for $\tau^+$ and $\tau^-$ decay products. This effect is well known for decay processes into a single neutral kaon and we do not discuss it much here. However, we mention some subtitles in section~\ref{sec:additional_effects}.

The asymmetry $A_{Exp,P}^{K}$ is the $CP$  violation due to kaon oscillation. We turn next to study it.

%For the $A_1$ asymmetry in Eq.~(\ref{A1_general}), 
%in the case of an ideal experiment with perfect efficiency $f_P(t)=1$, the asymmetry is given by Eq.~(\ref{A1_Bigi_Sanda}). In general, the measured asymmetry will have corrections  that depend on the efficiency function $f_1^K(t)$.  

%In the SM the first effect vanishes. In this work we only consider the SM.
%To understand the importance of the experimental cuts we define the experimental measure asymmetry 
%\begin{equation}\label{A1}
%A_{Exp,P}=\frac{\int_{0}^{\infty}dt\;[f_P(t)\Gamma(t) - \overline{f_P}(t)\overline{\Gamma}(t)]}{\int_{0}^{\infty}dt\;[f_P(t)\Gamma(t) + \overline{f_P}(t)\overline{\Gamma}(t)]}\;,
%\end{equation}
%to be the $CP$  asymmetry in the case where the neutral kaons $K^0,\;\overline{K}^0$ decay into $\pi\pi$. 
% Note that for precise theoretical predictions for experimental applications, we need to consider kaon oscillation in matter.\gp{Do we want this here?}

\subsection{Kaon decay rates}\label{subsec:kaon_decay_rates}
The asymmetry  $A_{Exp,P}^{K}$ defined in Eq.~(\ref{A_Exp}) can be computed using the decay rates of the kaon interaction eigenstates into two pions.
%, Eq.~(\ref{gamma_gammaBar}). 
For applications in a realistic experiment we need to take into account the effect of kaons oscillation in matter. Such effects were studied in the past and are also known as kaons regeneration in matter~\cite{Fetscher1996,Gsponer:1979,Ko:2011,ANGELOPOULOS1997422,Bjoern:2019qgw,Briere:1995,Quinn:2000,Uchiyama:1994, Roehrig}.
 
The kaon mass eigenstates are defined as 
\begin{equation} \ket{K_{S}}=p\ket{K^0} + q \ket{\overline{K}^0},\qquad 
\ket{K_{L}}=p\ket{K^0} - q \ket{\overline{K}^0},
\end{equation} 
with masses $m_S$ and $m_L$ and widths $\Gamma_S$ and $\Gamma_L$. Moreover, we define the experimentally known parameters:
\begin{align}
    m=\frac{m_S+m_L}{2},&\qquad \Gamma=\frac{\Gamma_S+\Gamma_L}{2},\\
    \Delta m = m_L - m_S,&\qquad \Delta\Gamma = \Gamma_L - \Gamma_S,
\end{align}
and 
\begin{equation}
A_{S}=\bra{\pi\pi}H\ket{K_{S}},\qquad A_{L}=\bra{\pi\pi}H\ket{K_{L}},
\end{equation}
as the decay amplitude for the mass eigenstates into two pions. We also define
\begin{equation}\label{xy}
x=\frac{\Delta m}{\Gamma},\qquad y=\frac{\Delta \Gamma}{2\Gamma}\;.
\end{equation}
In the case where direct $CP$  violation can be neglected, see, for example, Ref.~\cite{Grossman:2009}, we have
\begin{equation}
\frac{|p|^2-|q|^2}{|p|^2+|q|^2}= 2\text{Re}[\epsilon], \qquad \frac{\text{Im}[\epsilon]}{\text{Re}[\epsilon]}=-\frac{x}{y}.
\end{equation}
%where $\epsilon$ is the kaon $CP$  violating parameter.
The equations of motion for a neutral kaon state,
 \begin{equation}
\ket{\psi(t)}=c(t)\ket{K^0} + \overline{c}(t)\ket{\overline{K}^0} = c_S(t)\ket{K_S}+c_L(t)\ket{K_L},
 \end{equation}
are decoupled 
%in the base of massive kaon eigenstates $K_S$ and $K_L$, when the state is 
in vacuum:
\begin{equation}
c_S^{\text{vac}}(t) = c_S^0    \;e^{-im_St - \frac{\Gamma_S}{2}t}\;,\qquad c_L^{\text{vac}}(t) = c_L^0 \;e^{-im_Lt - \frac{\Gamma_L}{2}t}  ,
 \end{equation}
where $t$ is the time in the kaon frame, and $c_{S}^0$ and $c_{L}^0$ are the initial conditions of the time evolution. 

When a neutral kaon state propagates in matter, it can scatter by the strong interaction with the nuclei present in the material. In general, such scattering amplitude distinguishes between the interaction eigenstates $K^0$ and $\overline{K}^0$. This modifies the Hamiltonian of the system in basis of the mass eigenstates $K_S,K_L$ with an additional matter interaction term:
\begin{equation}
H_{\text{K,matter}}=\frac12\begin{pmatrix}
\chi+\overline{\chi}& ~\chi-\overline{\chi}\\
\chi-\overline{\chi} & ~\chi+\overline{\chi}
\end{pmatrix}\,,
\end{equation}
where $\chi,\overline{\chi}$ are defined in terms of the nuclear scattering amplitudes $\mathcal{A}$ and $\overline{\mathcal{A}}$ of the $K^0$ and $\overline{K}^0$ with matter \cite{Fetscher1996}:
\begin{equation}\label{chi_chiBar}
\chi = -\frac{2\pi n}{m_K}\mathcal{A}\;,\qquad \overline{\chi} = -\frac{2\pi n}{m_K}\overline{\mathcal{A}}\;,
\end{equation}
where $n$ is the scattering centers density, and $m_K$ is the neutral kaon mass (note that the amplitudes $\mathcal{A},\overline{\mathcal{A}}$ are commonly called $f,\overline{f}$ in the literature). We also define the so called kaons regeneration parameter:
\begin{equation}\label{r_def}
    r = \frac{1}{2}\frac{\chi-\overline{\chi}}{\Delta m - i\Delta\Gamma/2}\;.
\end{equation}
The value of $r$ is typically of the order of $10^{-2}$ or less~\cite{Bjoern:2019qgw,Fetscher1996,Ko:2011} and can be treated as a small parameter. Solving the equations of motion for a neutral kaon state propagating in matter and neglecting $\mathcal{O}(r^2)$ and $\mathcal{O}(r\epsilon)$ terms we find (see for example Ref.~\cite{Fetscher1996})
\begin{align}\label{cSL_matter}
    c_S(t) = e^{-\frac{i}{2}(\chi+\overline{\chi})t}\left[
c^0_S e^{-im_S t -\frac{\Gamma_S}{2} t} + r c_L^0(e^{-im_L t -\frac{\Gamma_L}{2} t} - e^{-im_S t -\frac{\Gamma_S}{2} t}) \right]\;,\nonumber \\
c_L(t) = e^{-\frac{i}{2}(\chi+\overline{\chi})t}\left[
c^0_L e^{-im_L t -\frac{\Gamma_L}{2} t} + r c_S^0(e^{-im_L t -\frac{\Gamma_L}{2} t} - e^{-im_S t -\frac{\Gamma_S}{2} t}) \right]\;.
\end{align}
Eq.~(\ref{cSL_matter}) can be used to compute the time evolution of the kaon interaction eigenstates $\ket{K^0(t)}$ and $\ket{\overline{K}^0(t)}$ created in the $\tau$ decay processes of our interest, and to find the kaon decay rates into two pions, that include regeneration effects. We find
\begin{align}\label{Gamma_Full}
    \Gamma_{\pi\pi}(t) = \frac{|A_S|^2}{4|p|^2}\left|e^{-\frac{i}{2}(\chi+\overline{\chi})t}\right|^2 \bigg[ 
 - 2\text{Re}[r]\left(e^{-\Gamma_s t}- e^{-\Gamma t}\left(\cos(\Delta m t) + \frac{\text{Im}[r]}{\text{Re}[r]}\sin(\Delta m t)\right)\right) \nonumber \\ 
+e^{-\Gamma_s t}- 2\text{Re}[\epsilon]e^{-\Gamma t}\left( \cos(\Delta m t) + \frac{\text{Im}[\epsilon]}{\text{Re}[\epsilon]}\sin(\Delta m t)\right)\bigg]\;,
\end{align}
and
\begin{align}\label{GammaBar_Full}
    \overline{\Gamma}_{\pi\pi}(t) = \frac{|A_S|^2}{4|q|^2}\left|e^{-\frac{i}{2}(\chi+\overline{\chi})t}\right|^2 \bigg[ 
 - 2\text{Re}[r]\left(-e^{-\Gamma_s t}+ e^{-\Gamma t}\left(\cos(\Delta m t) + \frac{\text{Im}[r]}{\text{Re}[r]}\sin(\Delta m t)\right)\right) \nonumber \\ 
+e^{-\Gamma_s t}+ 2\text{Re}[\epsilon]e^{-\Gamma t}\left( \cos(\Delta m t) + \frac{\text{Im}[\epsilon]}{\text{Re}[\epsilon]}\sin(\Delta m t)\right)\bigg]\;.
\end{align}
where we have used  $A_L/{A_S}\approx\epsilon$. As a consequence, we find that to leading order
\begin{equation}
    \Sigma_{\pi\pi}(t)=Ce^{-\Gamma_S t}\;,
\end{equation}
and that
\begin{align}\label{Delta_pi_pi_Full}
    \Delta_{\pi\pi}(t) = C\bigg[ \left(-2\text{Re}[\epsilon] - 2\text{Re}[r]\right)e^{-\Gamma_St} + 2\text{Re}[r]e^{-\Gamma t}\left( \cos(\Delta m t) + \frac{\text{Im}[r]}{\text{Re}[r]}\sin(\Delta m t) \right)\nonumber \\
    + 2\text{Re}[\epsilon]e^{-\Gamma t}\left( \cos(\Delta m t) - \frac{x}{y}\sin(\Delta m t) \right)\bigg]\;,
\end{align}
where we defined
\begin{equation}
    C = \frac{|A_S|^2}{4}\frac{|p|^2 + |q|^2}{|p|^2|q|^2}\left|e^{-\frac{i}{2}(\chi+\overline{\chi})t}\right|^2\,.
\end{equation}
Using Eq.~(\ref{chi_chiBar}) and the optical theorem, $\chi,\overline{\chi}$ can be related to the total interaction cross sections of $K^0$ and $\overline{K}^0$ with matter, see for example Ref.~\cite{Fetscher1996}.  For applications in realistic experiments, such cross sections are small, and the factor $\left|e^{-\frac{i}{2}(\chi+\overline{\chi})t}\right|^2\approx1$ can be ignored up to small corrections.
\begin{comment}
\begin{equation}
    \left|e^{-\frac{i}{2}(\chi+\overline{\chi})t}\right|^2 = e^{-\frac{n}{2}(\sigma + \overline{\sigma})\gamma\beta t}\;,
\end{equation}
where $\sigma$ and $\overline{\sigma}$ are the total cross sections of the interaction of $K^0$ and $\overline{K}^0$ with matter, $\beta$ is the velocity of the neutral kaon in the laboratory frame and $\gamma=\sqrt{1-\beta^2}$. For applications in realistic experiments, such cross sections are small, and the factor $\left|e^{-\frac{i}{2}(\chi+\overline{\chi})t}\right|^2\approx1$ up to small corrections that can be neglected. 
\end{comment}
At leading order in $\epsilon$ and $r$ we %find that
%\begin{align}\label{A_1_det}
%A_{Exp,1}^{\text{det}} = \frac{\int_0^{\infty}dt\;\Delta f_1^K(t)e^{-\Gamma_S t}}{\int_0^{\infty}dt\;f_1^K(t)e^{-\Gamma_S t}}\,,
%\end{align}
%and, separating the terms proportional to $\epsilon$ and to $r$, we get that 
can separate the asymmetry due to kaons oscillation into two contributions:
\begin{equation}\label{A_matter}
    A_{Exp,P}^K = A_{Exp,P}^{K,\text{vac}} + A_{Exp,P}^{K,\text{mat}}\;.
\end{equation}
The asymmetry $A_{Exp,P}^{K,\text{vac}}$ is due to kaons oscillation in vacuum, that is when $r=0$, and only depends on the $CP$  violating parameter $\epsilon$,
\begin{align}\label{A1_analytic}
    A_{Exp,1}^{K,\text{vac}} = -2\text{Re}[\epsilon]\;\left[1- \frac{\int_0^{\infty}dt\;f_1^K(t)e^{-\Gamma t}\left( \cos(\Delta m t) - \frac{x}{y}\sin(\Delta m t) \right) }{\int_0^{\infty}dt\;f_1^K(t)e^{-\Gamma_S t}}\right]\;.
\end{align}
The term $A_{Exp,P}^{K,\text{mat}}$, is the correction to the kaons oscillation asymmetry due to kaons regeneration in matter, and only depends on the parameter $r$, which depends on the properties of the material used in the experiment. We discuss this correction in Section~\ref{sec:additional_effects}.

For an ideal experiment, where $f_1^K(t)=1$, and the oscillation happens in vacuum, $r=0$, the integrals in Eq.~(\ref{A1_analytic}) can be performed analytically and we obtain
\begin{equation}\label{A1_ideal}
    A_{Ideal,1}^K = -2\text{Re}[\epsilon]\frac{\Gamma_S\left(x\Delta m\Gamma_S + y\Delta m^2 + y\Gamma^2 - y\Gamma\Gamma_S \right)}{y(\Delta m^2  + \Gamma^2) }\;.
\end{equation}
Using the definitions in Eq.~(\ref{xy}), and that from   $\Gamma_S\gg\Gamma_L$ we have $\Delta\Gamma\approx-\Gamma_S$, we can rewrite  Eq.~(\ref{A1_ideal}) in terms of $x$ and $y$ only:
\begin{equation}
    A_{Ideal,1}^K=-2\text{Re}[\epsilon]\frac{2y^2+yx^2-y}{yx^2+y}\approx +2\text{Re}[\epsilon]\;,
\end{equation}
where in the last step we have used  $x\approx -y$ and $x\approx 1$, as known from experimental kaons oscillation data. We note that while numerically the result agrees with that of \cite{Bigi:2005} this is just a numerical accident.

The generalization to the case of $A_2$ is straightforward and we get
\begin{align}\label{A_2_det}
    A_{Exp,2}^{\text{det}} = \frac{\int_0^{\infty}dt\;\Delta f_2^{\overline{K}}(t)e^{-\Gamma_S t}}{\int_0^{\infty}dt\;f_2^{\overline{K}}(t)e^{-\Gamma_S t}},
\end{align}
and 
\begin{align}\label{A2_analytic}
A_{Exp,2}^{K,\text{vac}} = +2\text{Re}[\epsilon]\;\left[1- \frac{\int_0^{\infty}dt\;f_2^{\overline{K}}(t)e^{-\Gamma t}\left( \cos(\Delta m t) - \frac{x}{y}\sin(\Delta m t) \right) }{\int_0^{\infty}dt\;f_2^{\overline{K}}(t)e^{-\Gamma_S t}}\right]\;,
\end{align}
which shows that in an ideal experiment in vacuum with perfect deficiency $f_1^{K}(t)=f_2^{\overline{K}}(t)=1$, we would have $A_2=-A_1$.

\section{The efficiency function}\label{sec:efficiency_function}

%In this section we discuss the efficiency function as it is central to the effect we are considering. 

The efficiency is a function of the kaon decay time $t$ in the kaon rest frame. As seen in Eq.~(\ref{f_P_ALL}) it depends on three indices. In this section, to simplify the notation, we write the efficiency function as $f_P^{\alpha}(t)$, where we define the multi-index $\alpha$ to be
\begin{equation}\label{alpha_index}
    \alpha= (+,K),\;(+,\overline{K}),\;(-,K),\;(-,\overline{K})\;,
\end{equation}
such that it defines whether we consider the efficiency for $K^0$ or $\overline{K}^0$, and whether the kaon is coming from $\tau^+$ or $\tau^-$ decay.
\begin{comment}
Even if the strategy that we present to obtain the function is general and works for any efficiency function, to simplify the notation, for the rest of this section, we will omit the superindices, and only keep the subscript $P$, that is $f_P^{^{...}}(t)=f_P(t)$.  However, we should keep in mind that the efficiency depends on the process $P$, as well as whatever the kaon is $K^0$ or $\overline{K}^0$, and whatever the kaon is coming from $\tau^+$ or $\tau^-$.
\end{comment}

\subsection{Obtaining the efficiency function}
The efficiency function fundamentally depends on
the kaon spectrum in the $\tau$ rest frame as well as the detector details and performance.

To understand how we obtain $f_P^{\alpha}(t)$, we first note that we have to consider three different frames: the kaon frame, the $\tau$ frame, and the laboratory (lab) frame. The fundamental two functions that define the efficiency function are:
\begin{itemize}
\item 
$F(L, E,\vec{p}, X_{\alpha})$: the laboratory efficiency function. It gives the reconstruction efficiency of the experiment for the detection of a neutral kaon, as a function of the kaon decay length $L$, energy $E$, momentum $\vec{p}$, and any other relevant parameters collected in the multivariable symbol $X_{\alpha}$, in the laboratory frame. %In general, this function may depend on whatever the kaon is a $K^0$ or a $\overline{K}^0$.
\item 
$S^{(\tau)}_{\alpha,P}(E^{(\tau)},\hat{n}^{(\tau)})$: the normalized kaon energy spectrum in the $\tau$ rest frame.
It is a function of the energy, $E^{(\tau)}$, and direction, $\hat{n}^{(\tau)}=\vec{p}^{\,(\tau)}/p^{(\tau)}$, of the kaon emitted in the process $P$ in the $\tau$ frame. 
%If there is a single neutral kaon in final state, and there is no direct $CP$  violation in the $\tau$ decay, then the kaon spectrum is the same for $\tau^+$ and $\tau_-$ decays.
\end{itemize}
We further define 
\begin{itemize}
\item 
$S_{\alpha,P}(E,\hat{n})$:  The normalized kaon energy spectrum in the laboratory frame. It is a function of the energy, $E$, and direction, $\hat{n}=\vec{p}/p$, of the kaon emitted in the process $P$ in the laboratory frame. %Due to effects like the forward-backward asymmetry in the $\tau^+\tau^-$ production, this spectrum can be different for $K^0$ and $\overline{K}^0$.
\end{itemize}

We obtain the laboratory frame spectrum, $S_{\alpha,P}$ from the spectrum in the $\tau$ frame, $S^{(\tau)}_{\alpha,P}$, by considering a boost from the $\tau$ frame to the laboratory frame:
\begin{equation}
    (E,\;\vec{p}\,) = \text{Boost}_{\tau\rightarrow\text{LAB}}(E^{(\tau)},\;\vec{p}^{\,(\tau)})\;.
\end{equation}
\begin{comment}
\begin{align}
    \begin{pmatrix}
E \\
\vec{p} 
\end{pmatrix} =
\begin{pmatrix}
\gamma_{\tau} & -\gamma_{\tau}\vec{\beta}_{\tau}^T \\
-\gamma_{\tau}\vec{\beta}_{\tau} & ~~~\delta_{ij}+(\gamma_{\tau}-1)\frac{\beta_{\tau,i}\beta_{\tau,j}}{\beta_{\tau}^2} 
\end{pmatrix}
\begin{pmatrix}
E^{(\tau)} \\
\vec{p}^{(\tau)} 
\end{pmatrix}
\end{align}
\end{comment}
%where $(E^{(\tau)},\vec{p}^{(\tau)})$ are the energy and momentum of the kaon in the $\tau$ frame, and $(E,\vec{p})$ in the laboratory frame. 
The boost depends on the direction and velocity of the $\tau$ in the laboratory frame. 

In general, the computation of the laboratory-frame spectrum $S_{\alpha,P}$ is nontrivial, since the velocity and angular distribution of the $\tau$ depend on both the center-of-mass energy and the asymmetry of the machine. While this is usually not an issue, the presence of neutrinos in the final state makes it problematic. Otherwise, one could simply combine all the daughter particles to reconstruct the $\tau$ momentum and direction. In practice, the spectrum must therefore be determined using numerical methods.

%We obtained the laboratory frame spectrum from the spectrum in the $\tau-$frame by boosting it. 

%$S^{(\tau)}_P(E^{(\tau)},\hat{n}^{(\tau)})$, by considering a boost from the $\tau$ frame to the laboratory frame:

%In general, the reconstruction efficiency in the laboratory frame  depends on the kaon's decay length, energy, momentum, whether the particle is a $K$ or $\Kbar$, and other aspects of the event. Thus, we define a laboratory frame efficiency function $F(L, E,\vec{p}, X_P)$, which gives the reconstruction efficiency of the experiment as a function of the decay length $L$, energy and momentum $E,\;\vec{p}$, and any other relevant parameters collected in the multivariable symbol $X_P$.

We express the efficiency as a function of the decay proper time in the kaon rest frame, $t$. The corresponding decay length in the laboratory frame can be written as $L=\beta\gamma t$, where $\beta$ denotes the kaon velocity in the laboratory frame and $\gamma=1/\sqrt{1-\beta^2}$ is the relativistic time dilatation factor. Both $\beta$ and $\gamma$ depend on the energy in the laboratory frame. Similarly, we can parametrize the additional parameters as $X_{\alpha}=X_{\alpha}(t)$. The total detection efficiency is then obtained  by averaging over the energy and momentum distribution of the kaon produced in process $P$ in the laboratory frame:
\begin{equation} \label{eq:genf}
    f_P^{\alpha}(t) = \int d\Omega\, dE \;S_{\alpha,P}(E,\hat{n})\;F(\beta\gamma t, E, \vec{p}, X_{\alpha}(t))\;,
\end{equation}
where $\Omega$ is the solid angle.

Eq.~(\ref{eq:genf}) is the main result of this section. It shows that the efficiency function $f_P^{\alpha}(t)$ factorizes into two components: the kaon spectrum $S_{\alpha,P}(E,\hat{n})$ in the laboratory frame, which depends on the specific process $P$; and a universal laboratory frame reconstruction efficiency $F(L,E,\vec{p},X_{\alpha})$, which depends only on the experiment conditions.  
%\ygn{I think it is a reuslt, in that we show that we can factorzed. Yet, we can rephare it, how about: Eq.~(\ref{eq:genf}) is important as it demonstrates that the ....}
To move forward we need to  understand the detector as well as the spectrum. Yet, many of the results can follow assuming a simplified version of both, as we turn to next.

%where $S_P(E,\hat{n})$ is the normalized spectrum of the kaon as a function of the energy $E$ and direction $\hat{n}=\vec{p}/p$ of the kaon emitted in the process $P$ in the laboratory frame. 
%We obtained the laboratory frame spectrum from the spectrum in the $\tau-$frame: $S^{(\tau)}_P(E^{(\tau)},\hat{n}^{(\tau)})$, by considering a boost from the $\tau$ frame to the laboratory frame:
%\begin{align}
%    \begin{pmatrix}
%E \\
%\vec{p} 
%\end{pmatrix} =
%\begin{pmatrix}
%\gamma_{\tau} & -\gamma_{\tau}\vec{\beta}_{\tau}^T \\
%-\gamma_{\tau}\vec{\beta}_{\tau} & \delta_{ij}+(\gamma_{\tau}-1)\frac{\beta_{\tau,i}\beta_{\tau,j}}{\beta_{\tau}^2} 
%\end{pmatrix}
%\begin{pmatrix}
%E^{(\tau)} \\
%\vec{p}^{(\tau)} 
%\end{pmatrix}
%\end{align}
%where $(E^{(\tau)},\vec{p}^{(\tau)})$ are the energy and momentum of the kaon in the $\tau-$frame, and $(E,\vec{p})$ in the laboratory frame. $\vec{\beta}_{\tau}$ is the velocity of the $\tau$ in the laboratory frame, and $\gamma_{\tau}=\sqrt{1-\beta_{\tau}^2}$. In general, the computation of the spectrum $S_P(E,\hat{n})$ can be non trivial, given that the velocity and angular distribution of the $\tau$ will depend on the detector and the asymmetry of the machine. It can be determined by monetacarlo simulation of the detector.

%\ygn{Not sure where to put it but i like to emphasis the factorization. that is that the result is the tau spectrum times boost time the detector function}\gp{I put an additional comment about this under Eq (\ref{f_P}), or you want more emphasis on the boost?}

\subsection{Simplification functions}
In this subsection we  make simplification approximations. We assume that $F$ is only a function of the kaon decay length and energy in the laboratory frame, $F=F(L,E)$. Since in this case $F$ does not depend on $\alpha$, it does not have any additional $CP$  violating effects. We can then use the kaon rest frame decay time, and the laboratory frame energy of the kaon to write it as  $F(\beta\gamma t,E)$. Under this assumption, the efficiency function is parameterized by the kaon decay time $t$ in its rest frame:
\begin{equation}\label{f_P}
    f_P^{\alpha}(t) = \int dE\;S_{\alpha,P}(E)F(\beta\gamma t,E).
\end{equation}
Eq.~(\ref{f_P}) shows that in this simplified case, the reconstruction efficiency $f^{\alpha}_P$ depends on the energy distribution of the kaon, which depends on the specific process and kaon in final state $(\alpha,P)$, and on $F(L,E)$, that is a universal experimental function, that does not depend on the process and on the the kaon detected in final state.

To illustrate the effect, we introduce a simple model that captures the essential features of the experiment. Specifically, we impose fixed decay-length cuts, $L_1$ and $L_2$, in the laboratory frame  to describe the finite geometrical acceptance of a particular detector, 
\begin{equation}\label{F_L}
    F(L,E)=F^{(L_1,L_2)}(L) = \theta\left(L-L_1\right)\theta\left(-L + L_2\right)\;,
\end{equation}
where $\theta$ is the step function. This represents a  simplified case in which  the laboratory frame efficiency function $F(L,E)$ depends solely on the decay length $L$.
For the above function, the time cuts in the kaon frame are energy-dependent. %If we fix $E$ to be the energy of the kaon, consistently with Eq.~(\ref{A_eps_t1_t2}), this makes the time cuts, and the thus asymmetry $A_{\epsilon}$ to be a function of such energy
%\begin{equation}
%    A^{(L_1,L_2)}_{\epsilon}(E) = A_{\epsilon}\left(\frac{L_1}{\beta\gamma},\frac{L_2}{\beta\gamma}\right)\;,
%\end{equation}
%where the energy dependence enters in the boost factor, $\beta \gamma$.
Given the efficiency $F(L,E)=F^{(L_1,L_2)}(L)$ as a simple geometrical acceptance model, we consider the case $P=1$ of the asymmetry $A_1$, where there is a single neutral kaon in the final state. Moreover, we assume that there is no direct $CP$  violation the $\tau$ decay, that the kaon oscillations happens in vacuum, $A_{Exp,1}^K=A_{Exp,1}^{1,\text{vac}}$, and we neglect a small effect by assuming that $\Delta f_1^K(t)=0$, which means that the spectrum $S_{1,K}^+(E)$ of the $K^0$ emitted by $\tau^+$, is the same as the spectrum $S_{1,\overline{K}}^-(E)$ of the $\overline{K}^0$ emitted by $\tau^-$. As a consequence, we can drop the index $\alpha$, and only consider $f_1^K(t)=f_1(t)$, see Eq.~(\ref{f_P_FULL_FULL}).
Under these assumptions, the experimentally accessible   asymmetry $A_{Exp,1}=A_{Exp,1}^{K,\text{vac}}=A_{Exp,1}^{(L_1,L_2)}$ is given by Eq.~(\ref{A_Exp}), where the kaon rest frame efficiency function takes the form
\begin{equation}\label{A1_experimental}
    f_{1}^{(L_1,L_2)}(t) = \int S_{1}(E)\;\theta\left(t-\frac{L_1}{\beta\gamma}\right)\theta\left(-t + \frac{L_2}{\beta\gamma}\right)dE\,.
\end{equation}

To study the impact  of the efficiency function on the experimental asymmetry, and its dependence on the kaons energy, we consider a simplified toy model in which the kaons are emitted with a fixed laboratory-frame energy $E_0$. In this case the spectrum reduces to
\begin{equation}
S_{1} = \delta(E-E_0).
\end{equation}
The efficiency function gives fixed-time experimental cuts that depends on the energy,
\begin{equation}
f_{1}^{(L_1,L_2)}(t) = \theta\left(t-\frac{L_1}{\beta\gamma}\right)\theta\left(-t + \frac{L_2}{\beta\gamma}\right)\,,
\end{equation}
where $\beta\gamma=\sqrt{E_0^2-m_K^2}/m_K$. Using Eq.~(\ref{A_Exp}), the asymmetry $A_{Exp,1}^{(L_1,L_2)}(E_0)$ modifies as
\begin{equation}\label{A1L1L2E_proof}
    A_{Exp,1}^{(L_1,L_2)}(E_0) = \frac{\int_{L_1/\beta\gamma}^{L_2/\beta\gamma}dt\;[\Gamma_{\pi\pi}(t) - \overline{\Gamma}_{\pi\pi}(t)]}{\int_{L_1/\beta\gamma}^{L_2/\beta\gamma}dt\;[\Gamma_{\pi\pi}(t) +\overline{\Gamma}_{\pi\pi}(t)]}\;.
\end{equation}
As illustrated by eq.~(\ref{A1L1L2E_proof}), the kaons emitted with different energies, have different efficiency functions, and, as a consequence, a different measured asymmetry $A_{Exp,1}^{(L_1,L_2)}(E_0)$. Using Eq.~(\ref{A1_analytic}), the asymmetry can be written as
\begin{align}\label{A_eps_t1_t2}
A^{(L_1,L_2)}_{Exp,1}(E_0)=
%\nonumber \\
-2\text{Re}[\epsilon]\left(1- \frac{I_{\rm int} }{I_{\rm tot}}\right)\;,
\end{align}
%where $\epsilon$ is the standard $CP$ violation parameter of the kaon system,
with
\begin{align}    
I_{\rm int}=  &\int_{L_1/\beta\gamma}^{L_2/\beta\gamma}dt\,\exp(-\Gamma t)\left[\cos(\Delta m t)  -\frac{x}{y}\sin(\Delta m t)\right],\\
I_{\rm tot} = &\int_{L_1/\beta\gamma}^{L_2/\beta\gamma} dt \,\exp\left(-\Gamma_S t\right).
\end{align}

\begin{figure}
\centering
\includegraphics[width=0.8\columnwidth]{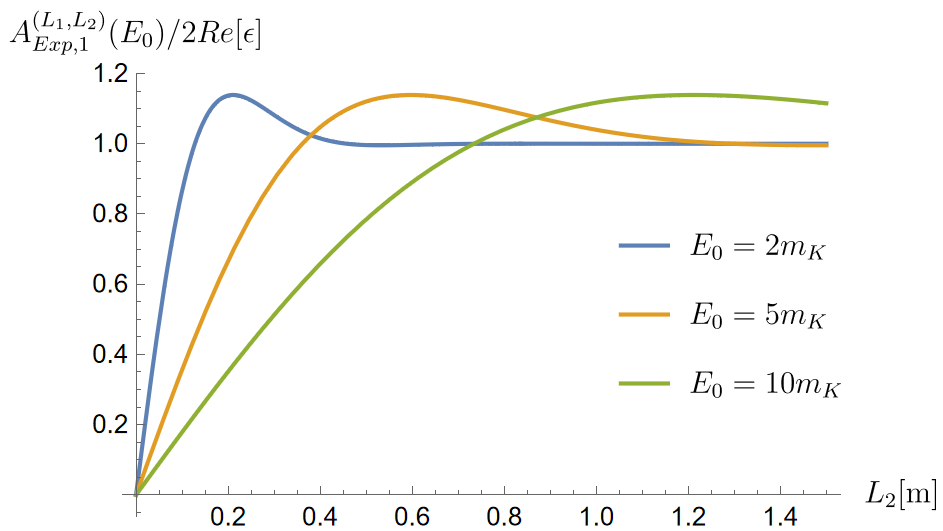}
\caption{The asymmetry $A_{Exp,1}^{(L_1,L_2)}(E_0)$ as a function of the total length of the experiment $L_2$ for different values of the kaon energy $E_0$. In this plot we set $L_1=\tau_Sc /10\approx 3\text{mm}$,  noting that for Belle~II the average energy of the emitted kaons is $\langle E\rangle\approx 5m_{K^0}$, with tails of the spectrum extending up to approximately $10m_{K^0}$.}
\label{fig:L2_plot}
\end{figure}

\begin{figure}
\centering
\includegraphics[width=0.8\columnwidth]{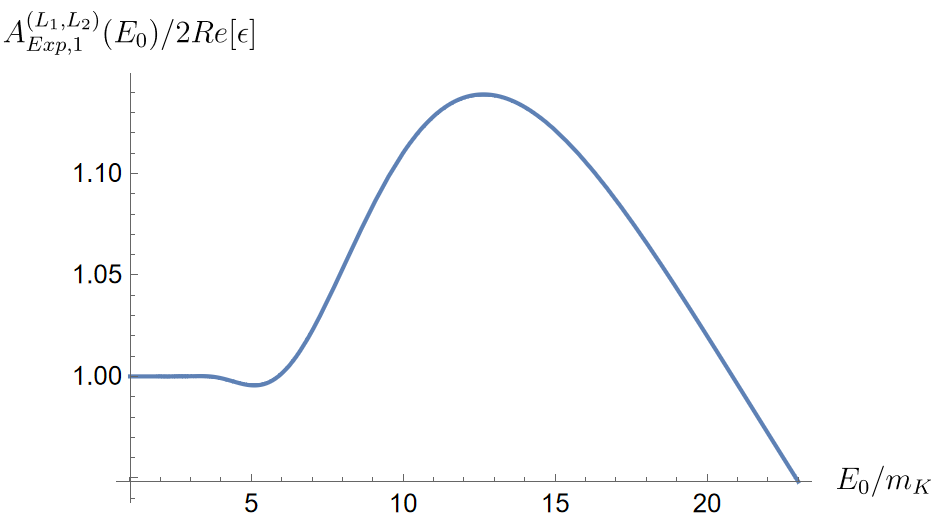}
\caption{The asymmetry $A_{Exp,1}^{(L_1,L_2)}(E_0)$ as a function of the kaon energy $E_0$. In this plot we set $L_1=\tau_Sc /10\approx 3\text{mm}$ and $L_2=\tau_Lc /10\approx 1.5\text{m}$.}
\label{fig:L_E_plot}
\end{figure}

Fig.~\ref{fig:L2_plot} displays the asymmetry $A_{Exp,1}^{(L_1,L_2)}(E_0)$ for different fixed values of the  kaon energy $E_0$ in the laboratory frame, as a function of the decay length $L_2$ of the experiment. % and for different values of the energy. 
In contrast, Fig.~\ref{fig:L_E_plot} shows the asymmetry $A_{Exp,1}^{(L_1,L_2)}(E_0)$ as a function of $E_0$ for fixed values of $L_1$ and $L_2$.

%%%%%%%%%%%%%%%%%%%%
\section{$CP$ violation in decays involving two neutral kaons}\label{sec:A3}

We now consider the case of the $A_3$ asymmetry, Eq.~(\ref{A3}). In terms of the kaon interaction eigenstates created, we are looking at the process $\tau\rightarrow\pi K^0 \overline{K}^0 \nu$, where the $K^0$ and $\overline{K}^0$ pairs are identified as $K_S$ and $K_L$ pairs. As before, we consider the case where the $K_S$  is identified by a $\pi\pi$ system in the final state, while the $K_L$ is identified as missing in the final detection due to its long lifetime. %This makes the $\tau\rightarrow\pi K^0 \overline{K}^0 \nu$ process as a background for the $A_1$ measurement.
This process is particularly relevant because it contributes as a background to the $A_1$ measurement. 
%Note that already by calling a state that decay into two pions a $K_S$ we are making an important approximation of order $\epsilon$. Yet, as long as it is clear what we are measuring that abuse of language is not a problem. 
The difference in this case compared to  $A_1$ and $A_2$ discussed earlier is that, because there are two neutral kaons, $K^0$ and $\overline{K}^0$, in the final state, we have to consider four possible different efficiency functions, denoted as $f_P^{+,K}(t)$, $f_P^{-,K}(t)$, $f_P^{+,\overline{K}}(t)$, $f_P^{-,\overline{K}}(t)$. In general, the difference between the $K^0$ and $\overline{K}^0$ efficiency functions makes the $CP$ asymmetry ratio, Eq.~(\ref{A3}), non-vanishing. 

For an ideal experiment without cuts and perfect efficiency, $A_3$ is zero. The reason is that, since both $K^0$ and $\Kbar$ are produced, the probability to have a $\pi\pi$ system in the final state  is the same for $\tau^+$ and  $\tau^-$ decay. 
In realistic experiments, however, 
finite efficiency introduces artificial asymmetry. In particular, this arises because the efficiency functions $f_P^K(t)$ and $f_P^{\overline{K}}(t)$ depend on the kaons spectrum. The kaon interaction eigenstates $K^0$ and $\Kbar$ are produced with a different energy spectra in $\tau$ decay. Thus, a more energetic $K^0$ has a different probability of arising from a $\tau^+$ compared to a $\tau^-$. As a consequence, the experimental cuts can bias the number of $K_S$ originating 
%a difference in the sample of $K^0$ and $\Kbar$ coming 
from $\tau^+$ and $\tau^-$. %The $K_S$ has different probability to emerge from a $K^0$ and $\Kbar$. In other words, since $K_S$ is defined by a $\pi\pi$ system, $K^0$ and $\Kbar$ have a different probability to decay into two pions. 
%Then, the cuts create a biased on the amount of the $K_S$ originating from the decay of the $\tau^+$ and the $\tau^-$.  
 
To make this explicit, we consider the case where $K^0$ and $\Kbar$ are emitted with a different energy spectra in the $\tau^{\pm}\rightarrow\nu K^0 \overline{K}^0 \pi^{\pm}$ process. 
We define $S_{K}^{\pm}(E)$ and $S_{\overline{K}}^{\pm}(E)$ as the energy spectra of the $K^0$ and the $\overline{K}^0$ emitted in the decay of the $\tau^{\pm}$ in the laboratory frame. 
Note that, to simplify the notation, the subscript $P=3$ is implicit. The experimentally measured asymmetry $A_3$  is then given by
\begin{align}\label{A_3_long}
A_{Exp,3}=\frac{\int_{0}^{\infty}dt\;\left[\left(f_3^{+,K}(t)\Gamma_{\pi\pi}(t) + f_3^{+,\overline{K}}(t)\overline{\Gamma}_{\pi\pi}(t)\right) - \left(f_3^{-,K}(t)\Gamma_{\pi\pi}(t) + f_3^{-,\overline{K}}(t)\overline{\Gamma}_{\pi\pi}(t)\right)\right]}{\int_{0}^{\infty}dt\;\left[\left(f_3^{+,K}(t)\Gamma_{\pi\pi}(t) + f_3^{+,\overline{K}}(t)\overline{\Gamma}_{\pi\pi}(t)\right) + \left(f_3^{-,K}(t)\Gamma_{\pi\pi}(t) + f_3^{-,\overline{K}}(t)\overline{\Gamma}_{\pi\pi}(t)\right)\right]}\;.
\end{align}
Assuming that direct $CP$ violation in  $\tau$ decays is negligible, we can write
\begin{equation}\label{CPV_direct}
S_{K}^+(E)= S_{\overline{K}}^-(E), \qquad
S_{K}^-(E)= S_{\overline{K}}^+(E).
\end{equation}
Thus, in the following we omit the $+$ superindex and make it implicit:
\begin{equation}\label{S_3_K}
S_{K}= S_{K}^+ ,\qquad S_{\overline{K}} = S_{\overline{K}}^+.
\end{equation}

Note that while Eq.~(\ref{CPV_direct}) is exact  in the $\tau$ rest frame,  small deviations may arise in the laboratory frame due to effects such as the forward-backward asymmetry in the $\tau^+\tau^-$ production (see  Section~\ref{sec:additional_effects} ). These corrections, however, are subleading and can be  neglected in the present  analysis.

From the efficiency function Eq.~(\ref{eq:genf}), we then obtain 
\begin{equation}\label{A3_f_CP_direct}
    f_3^{+,K}(t) = f_3^{-,\overline{K}}(t)=f_3^K(t)\;,\qquad f_3^{+,\overline{K}}(t) = f_3^{-,K}(t)=f_3^{\overline{K}}(t)\;.
\end{equation}
As a consequence, only the two efficiency functions $f_3^{K}(t)$ and $f_3^{\overline{K}}(t)$ defined in Eq.~(\ref{f_sum}) are  relevant.

To demonstrate the effect, we consider the simplified case where the reconstruction efficiency of the detector in the laboratory frame is given by $F(L,E)$. Then, using Eq.~(\ref{f_P}), the kaon frame efficiency function for the $K^0$ and $\overline{K}^0$ emitted by the $\tau^{+}$ are
\begin{align}\label{f_3_K}
    f_3^{K} (t) = \int dE\; S_{K}(E) F(\beta\gamma t,E), \qquad
    f_3^{\overline{K}} (t) = \int dE\; S_{\overline{K}}(E) F(\beta\gamma t,E)\;.
\end{align}
 Using Eq.~(\ref{A3_f_CP_direct}), the asymmetry  can be rewritten as
\begin{align}
A_{Exp,3}=\frac{\int_{0}^{\infty}dt\;\left[\left(f_3^{K}(t)\Gamma_{\pi\pi}(t) + f_3^{\overline{K}}(t)\overline{\Gamma}_{\pi\pi}(t)\right) - \left(f_3^{\overline{K}}(t)\Gamma_{\pi\pi}(t) + f_3^{K}(t)\overline{\Gamma}_{\pi\pi}(t)\right)\right]}{\int_{0}^{\infty}dt\;\left[\left(f_3^{K}(t)\Gamma_{\pi\pi}(t) + f_3^{\overline{K}}(t)\overline{\Gamma}_{\pi\pi}(t)\right) + \left(f_3^{\overline{K}}(t)\Gamma_{\pi\pi}(t) + f_3^{K}(t)\overline{\Gamma}_{\pi\pi}(t)\right)\right]}\;,
\end{align}
and using Eq.~(\ref{Delta_Sigma_pi_pi}),
\begin{equation}\label{A3_proof_final}
A_{Exp,3}=\frac{\int_0^{\infty}dt\;f_3^K(t)\Delta_{\pi\pi}(t) - \int_0^{\infty}dt\;f_3^{\overline{K}}(t)\Delta_{\pi\pi}(t)}{\int_0^{\infty}dt\;f_3^K(t)\Sigma_{\pi\pi}(t) + \int_0^{\infty}dt\;f_3^{\overline{K}}(t)\Sigma_{\pi\pi}(t)}\;.
\end{equation}
This expression can be rewritten in a more convenient form as
\begin{comment}
\begin{align}
A_{Exp,3}=\frac{\int_{0}^{\infty}dt\;\left[f_3^{K}(t)\left(\Gamma_{\pi\pi}(t) -\overline{\Gamma}_{\pi\pi}(t)\right) - f_3^{\overline{K}}(t)\left(\Gamma_{\pi\pi}(t) - \overline{\Gamma}_{\pi\pi}(t)\right)\right]}{\int_{0}^{\infty}dt\;\left[f_3^{K}(t)\left(\Gamma_{\pi\pi}(t) -\overline{\Gamma}_{\pi\pi}(t)\right) - f_3^{\overline{K}}(t)\left(\Gamma_{\pi\pi}(t) - \overline{\Gamma}_{\pi\pi}(t)\right)\right]}\;.
\end{align}
We note that the first and the second terms at the numerator recall the experimental measure asymmetry for a single kaon in final state $A_{Exp,P}$ given in Eq.~(\ref{A_Exp}), for two different efficiency functions $f_3^K(t)$ and $f_3^{\overline{K}}(t)$ respectively,
\begin{equation}\label{A3_proof_final}
A_{Exp,3}=\frac{\int_0^{\infty}dt\;f_3^K(t)\Delta_{\pi\pi}(t) - \int_0^{\infty}dt\;f_3^{\overline{K}}(t)\Delta_{\pi\pi}(t)}{\int_0^{\infty}dt\;f_3^K(t)\Sigma_{\pi\pi}(t) + \int_0^{\infty}dt\;f_3^{\overline{K}}(t)\Sigma_{\pi\pi}(t)}\;,
\end{equation}
which can be rewritten as}
\end{comment}
\begin{equation}\label{A3_Final}
    A_{Exp,3} = \frac{\int_0^{\infty}dt\;\Delta f_3(t)\Delta_{\pi\pi}(t)}{\int_0^{\infty}dt\; f_3(t)\Sigma_{\pi\pi}(t)},
\end{equation}
where the efficiency functions $f_P(t)$ and $\Delta f_P(t)$ with no superindex are defined in Eq.~(\ref{f_P_FULL_FULL}). 
The term $\Delta f_3(t)$  quantifies the difference between the efficiency functions for $K^0$ and  $\overline{K}^0$, reflecting their  different energy spectra in the $\tau^+$ decay, see Eq.~(\ref{f_3_K}). This contrasts with  $A_{Exp,1}^{\text{det}}$ defined in Eq.~(\ref{A1_Final}), where the efficiency differences  between two $CP$ conjugates modes arise from  detector-related effects.

We emphasize that $A_3\neq0$ originates from the different energy spectra of $K^0$ and $\overline{K}^0$  in the $\tau$ rest frame, which translates into different spectra also in the laboratory frame: $S_{K}(E)\neq S_{\overline{K}}(E)$.
%in the decay of the $\tau$. 
Consequently, the dependence of $f_P(t)$ on the energy spectrum, see Eq.~(\ref{f_P}), prevents cancellation of asymmetries.
%As a consequence, the dependence of the rest frame efficiency $f_P(t)$ on the energy spectrum, Eq.~(\ref{f_P}), makes the asymmetries due to $K^0$ and $\overline{K}^0$ not cancel. 
Indeed, if $S_{K}=S_{\overline{K}}$, then  $\Delta f(t)=0$, and  $A_3$ would vanish. In contrast, $A_1$ and $A_2$ correspond to genuine physical asymmetries that would exist even with perfect detection efficiency $F(L,E)=f_P(t)\equiv1$. By comparison, $A_3$ arises solely  from the limited reconstruction efficiency of the experiment.

%%%%%%%%%%%%%%%%%%%
\section{Relevance for Belle~II}\label{sec:BELLE2}

In this section we discuss the SM predictions for the asymmetries $A_1$, $A_2$, and $A_3$ relevant to a Belle~II-type experiment~\cite{BELLE2:2018}. We emphasize that this constitutes a simplifying analysis and a complete analysis must be carried out in conjunction with the experimental data. As shown in Eq.~(\ref{eq:genf}), the calculation requires knowledge of the detector efficiency function $F$ as well as the spectra of $K^0$ and $\overline{K}^0$ mesons produced in  $\tau$ decays across the different processes.

%Since the $K_S$ in final state is identified as a $\pi\pi$ system, the asymmetry Eq.~(\ref{A3}) is not experimentally distinguishable from Eq.~(\ref{A1}). We aim to give an estimate of this background effect with applications for Belle~II experiment. To do so we need the information of the spectrum of $K^0$ and $\overline{K}^0$ emitted in the decay of the $\tau$. 

\subsection{Belle~II Efficiency Function}\label{subsec:BELLE2_simulation}
To compute the reconstruction efficiency function for a Belle~II-type experiment, we adopt the simplified model of Eq.~(\ref{f_P}), in which the laboratory-frame efficiency function $F$ is assumed to depend only on the decay length $L$ and the energy $E$ of the detected kaon, $F=F(L,E)$. 
 
We generate $10^7$ $K_S$ mesons under Belle~II kinematics constraints. For each $K_S$, we simulate the decay length $L$, the decay time $t$ in the kaon rest frame, and the laboratory-frame energy $E$. 
The full sample is then divided into equipopulous 10 bins of laboratory-frame energy $\Delta E_i$. For each bin $\Delta E_i$, we construct the decay-time distribution. 
%We then use an  approximate model to simulate the Belle~II detector response for $\tau$ decays into kaons.
These distributions are subsequently smeared to mimic the resolution effects, providing a set of reference distributions, denoted $D^0_i(t)$.  
In this approximate model, the detector efficiency $F$ is parameterized as %across the full spectrum of the events observed in the experiment:
\begin{equation}\label{F_BELLE2}
    F(L,E) = F_i(L) \;\text{for}\;E\in\Delta E_i=[E_i,E_{i+1}]\;\;\;i=1,...,10\;.
\end{equation}
%In this way, we can use Eq. (\ref{f_P}) to compute the total efficiency function for a given process as 
%\begin{equation}\label{f_P_BELLE2_general}
%    f_P^{\alpha}(t) = \sum_{i=1}^{10} \int_{\Delta E_i} dE\; S_{\alpha,P}(E)F_i(\beta\gamma t)\;,
%\end{equation}
\begin{comment}
\textcolor{blue}{[To be edited by Paolo,...]: Some first info about the efficiency function calculation can be added in this part of the section, the rest will be explained better in appendix. Here we can use few words to describe the strategy to compute the efficiency function for different values of the energy, and mention we used Geant4 to simulate the Belle~II apparatus}. 
\end{comment}
%We performed an independent study without using the Geant4 Belle~II detector simulation.
%
%We used the following steps as the strategy to extract the efficiency function $F$, Eq.~(\ref{F_BELLE2}).
%We generate $10^7$ $K_S$ with Belle~II kinematics constraints, and for each $K_S$ we simulate the decay length $L$, decay time in the kaon frame $t$, and the laboratory frame energy $E$. 
%We then divide the total sample of events into 10 bins of different laboratory frame energies $\Delta E_i$, such that each bin has the same number of events.
%For each energy bin $\Delta E_i$, we built the decay time distribution. We then smear such distributions to mimic the resolution effects. In this way we obtain a reference generated distribution that we call $D^0_i(t)$.
%
To each decay-time distribution $D_i^0(t)$, we apply experimental cuts given by the acceptance of the detector and by selection analysis typical of a Belle~II like experiment:
\begin{align}
    \begin{cases}
        0.3 < \theta(\pi^{\pm})[\text{rad}]<2.62\;,\\
        0.2 < p_T(K_S)[\text{GeV}]<4.5\;,\\
        p_T(\pi^{\pm})[\text{GeV}]>0.1\;,\\
        K_S\;\text{decay length}[\text{cm}]>0.2
    \end{cases}
\end{align}
where $\theta$ is the angle with respect to the beam pipe, and $p_T=p\sin\theta$ is the transverse momentum. Moreover, in order to reproduce the loss of efficiency for higher decay times of the $K_S$, we model the vertex fitting efficiency as a linear function of the flight distance in the range $[0.2,\;120]\,\text{cm}$.
This yields a new set of distributions $D^{exp}_i(t)$, which mimic those expected in experimental data.

\begin{figure}
%\centering
\includegraphics[width=0.8\columnwidth]{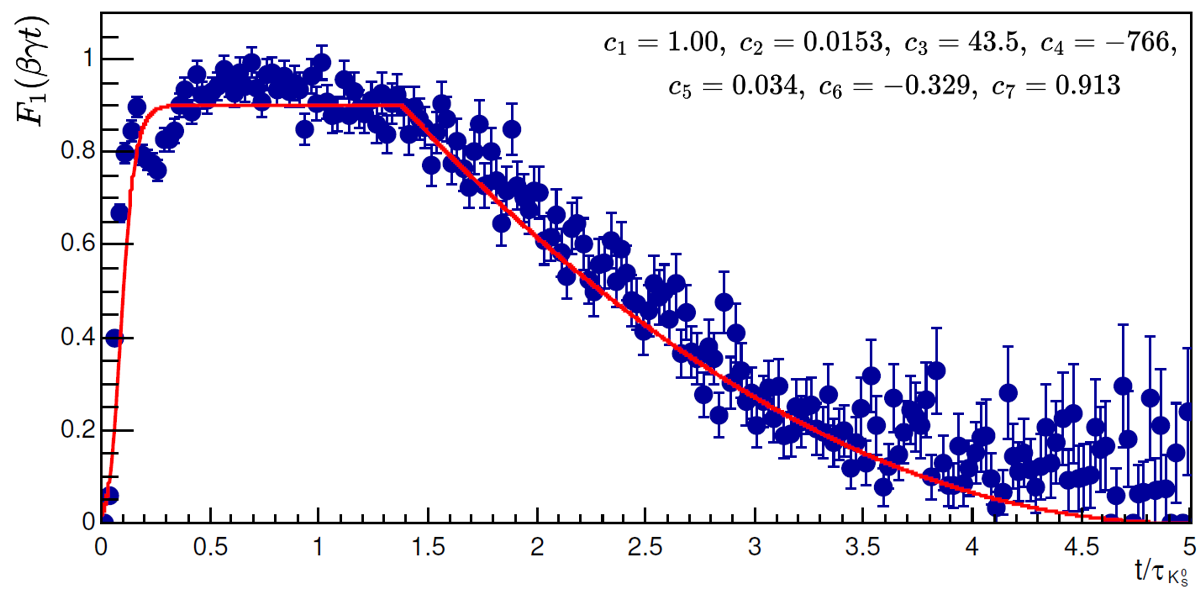}
\caption{Example of the approximate detector efficiency model for a Belle~II-type experiment for the first energy bin $\Delta E_1=[0.55,1.23]\;\text{GeV}$. The red curve is the result of the fit using Eq.~(\ref{FIT_BELLE2}), with the fit parameters showed in the upper right of the figure.}
%$c_1=1.00$, $c_2=0.0153$, $c_3=43.5$, $c_4=-766$, $c_5=0.034$, $c_6=-0.329$, $c_7=0.913$.
\label{fig:eff_BELLE2}
\end{figure}

The efficiency in each bin is then extracted as a function of decay length, expressed in terms of the boosted decay time $L=\beta\gamma t$, by measuring the ratio:
\begin{equation}\label{Dexp/D0}
F_i(\beta\gamma t)=\frac{D^{exp}_i(t)}{D^0_i(t)},\;\text{for}\;E\in\Delta E_i\;.
\end{equation}
To obtain an analytic expression of $F_i(\beta\gamma t)$, we  fit the above ratio  with the following seven-parameter function:
\begin{equation}\label{FIT_BELLE2}
 F_i(\beta\gamma t) =\begin{cases}
(c_1 + c_2 e^{-c_3 t})^{c_4}\qquad\text{for}\;t\in[0,t_1]\,, \\
c_5 t^2 + c_6 t + c_7 \qquad\text{for}\;t\in[t_1,t_2]\,,
\end{cases}
\end{equation}
where $c_1,...,c_7$ are the fit parameters, each  implicitly dependent on $i=1,...,10$. The transition point  $t_1$ is determined separately for every energy range $\Delta E_i$ to give the best fit, while the upper bound is fixed at $t_2=6\;\tau_S$ for all generated events. Finally, the results across all bins are combined to construct a global model for the efficiency function $F(\beta\gamma t)$. 
%We then combined all of the bins to generate a model for $F(\beta\gamma t)$.
%
Fig.~\ref{fig:eff_BELLE2} illustrates an example of the detector efficiency obtained for the first energy bin $\Delta E_1$, together with the corresponding fit. The behavior in other energy bins is similar.

\begin{comment}
In Fig.~\ref{fig:sub1}, we show the effect that the different efficiency functions $F_i(\beta\gamma t)$ have for different values of the energy of the kaons. In particular, the plot shows the value of the experimental asymmetry $A_{Exp,1}^K$, Eq. (\ref{A1_Final}), computed for different efficiency functions $f_{1,i}^K(t)=F_i(\beta\gamma t)$, that is without performing the average in the total spectrum Eq.~(\ref{f_P}). Fig.~\ref{fig:sub2} shows the total spectrum of the kaons generated in the simulation of the Belle~II detector.
\end{comment}

\begin{comment}
\begin{figure}
\centering
\begin{subfigure}{.5\textwidth}
  \centering
  \includegraphics[width=\linewidth]{A1_E.png}
  \caption{}
  \label{fig:sub1}
\end{subfigure}%
\begin{subfigure}{.5\textwidth}
  \centering
  \includegraphics[width=\linewidth]{Energy_distribution.png}
  \caption{}
  \label{fig:sub2}
\end{subfigure}
\caption{On the left, we show the asymmetry $A_{Exp,1}^K$, Eq. (\ref{A1_Final}), in  the case that $f_1^K(t)=F_i(\beta\gamma t)$ where $F_i$ is computed for the different energy intervals $\Delta E_i$. On the right, we show the normalized energy spectrum of the kaons $K_S$ generated in the simulation of the Belle~II detector.}
\label{fig:test}
\end{figure}
\end{comment}

\subsection{Predictions for $A_1$ and $A_2$ }\label{subsec:BELLEII_A1_A2}
%We aim to give a theoretical prediction for the experimental asymmetries $A_{Exp,1}$, Eq.~(\ref{A1_Final}) and $A_{Exp,2}$, Eq.~(\ref{A2_Final}), in a Belle~II type experiment. 
As shown in Eq.~(\ref{A1_Final}) and Eq.~(\ref{A2_Final}), the experimental asymmetries $A_{Exp,1}$ and $A_{Exp,2}$ depend on 
the efficiency function $f^{\alpha}_P(t)$, which in turn depends on the kaon energy distribution of the specific decay. Since the energy spectrum of the neutral kaon  in $\tau^{\pm}\rightarrow\nu K^{\pm}K_S$ differs from that in $\tau^{\pm}\rightarrow\nu\pi^{\pm}K_S$, one generally expects $A_1\neq -A_2$  in real experimental measurements.

In the $U$-spin limit, however, the two spectra relevant for $A_1$ and $A_2$ become identical, leading to $A_1 = -A_2$. Deviations from this relation arise from $U$-spin breaking effects, which are expected to be small.
%We note that the in the U-spin limit the two spectra for $A_1$ and $A_2$ are the same and thus we do have $A_1=-A_2$. The effect of U-spin breaking is expected to be small.

To determine the asymmetries, we first  compute the efficiency functions $f_1^{K}(t)$ and $f_2^{\overline{K}}(t)$. For simplicity, we assume that $\Delta f_1^K=0$ and $\Delta f_2^{\overline{K}}=0$, thereby neglecting detector-related systematic effects. Furthermore, we identify the experimental asymmetries with their vacuum counterparts, $A_{Exp,1}=A_{Exp,1}^{K,\text{vac}}$ and 
that $A_{Exp,2}=A_{Exp,2}^{K,\text{vac}}$, which amounts to neglecting  matter-interactions effects.  The detector efficiency is modeled according to Eq.~(\ref{F_BELLE2}), and the total efficiency functions are obtained through  Eq.~(\ref{f_P}). For the case of $A_1$, we require  the laboratory frame $K^0$ energy spectrum $S_{K,1}(E)$ associated with the decay $\tau^+\rightarrow \overline{\nu}\pi^+ K^0$ process:
\begin{equation}
    f_1^K(t) = \sum_{i=1}^{10} \int_{\Delta E_i} dE\; S_{K,1}(E)F_i(\beta\gamma t)\;.
\end{equation}
Similarly, for $A_2$, the relevant input is the laboratory frame $\overline{K}^0$ spectrum $S_{\overline{K},2}(E)$ from the decay the $\tau^+\rightarrow \overline{\nu}K^+ \overline{K}^0$ process:
\begin{equation}
    f_2^{\overline{K}}(t) = \sum_{i=1}^{10} \int_{\Delta E_i} dE\; S_{\overline{K},2}(E)F_i(\beta\gamma t)\;.
\end{equation}

We first compute the energy spectra $S^{(\tau)}_{K,1}$ and $S^{(\tau)}_{\overline{K},2}$ in the $\tau$ rest frame. For this purpose,  we adopt a simplified model of a point-like  three-body decay. In this approximation, the difference between the two spectra arises from the available kinematic phase space.  This treatment provides only a rough estimate, sufficient to gauge the impact of kaon energy distributions on the experimental asymmetries. Further details of the spectrum computation in the $\tau$ rest frame are given in Appendix~\ref{app:Kaon_spetrum}.
%This is just a rough approximation that we use to have an estimate of the effects of the kaons energy distributions on the experimental asymmetries. For more details on the $\tau$ frame spectrum computation, see Appendix~\ref{app:Kaon_spetrum}. 

To obtain the laboratory-frame spectra, $S_{K,1}(E)$ and $S_{\overline{K},2}(E)$, we boost the results from the $\tau$ rest frame. 
%Then, we make a boost to the laboratory frame to find $S_{K,1}(E)$ and $S_{\overline{K},2}(E)$.
For Belle~II-type experiment,  we assume  an asymmetric electron-positron collider operating at an  invariant mass $\sqrt{s}=10.58\;\text{GeV}$, with the entire center-of-mass energy used for $\tau^+\tau^-$ production. The collider asymmetry originates from the unequal beam energies, taken here as $4\, \text{GeV}$ and $7\,\text{GeV}$. 
The angular distribution of the $\tau^+\tau^-$ pairs in the center-of-mass frame is determined from the $e^+e^-\rightarrow \tau^+\tau^-$ cross section. 
%The collider asymmetry is defined by the different energies of the colliding particles: $4\;\text{GeV}$ and $7\;\text{GeV}$.  

Using the analytic expressions in Eq.~(\ref{A1_analytic}) and Eq.~(\ref{A2_analytic}), we obtain numerical estimates for the asymmetries:
\begin{align}
A_{Exp,1}& \approx 3.00\times 10^{-3}\; \approx 0.909\times 2\text{Re}[\epsilon]\,, \\
A_{Exp,2} &\approx -3.07\times 10^{-3}\; \approx -0.931\times 2\text{Re}[\epsilon]\,.
\end{align}
Thus, for a Belle~II-type experiment the relation $A_2=-A_1$ holds to within an error of order:
\begin{equation}\label{A12_result}
    \frac{A_{Exp,1}+A_{Exp,2}}{2\text{Re}[\epsilon]}\approx 2\%\;.
\end{equation}
While we do not assign an  error on the above estimate, we do expect that the result is a correct rough estimate and thus that the effect is within the range of a few percent. Thus, we conclude that 
this deviation is expected to remain below the experimental sensitivity.

\subsection{Prediction for $A_3$ }
We now turn to a theoretical estimate of the experimental asymmetry $A_{Exp,3}$, defined in  Eq.~(\ref{A3_Final}). 
This requires the evaluation of the efficiency functions $f_3^K(t)$ and $f_3^{\overline{K}}(t)$ from Eq.~(\ref{f_3_K}), and consequently $\Delta f_3(t)$. Hence, we need a theoretical model for  the kaon energy distribution functions $S_{K}(E)$ and $S_{\overline{K}}(E)$ in the $\tau^+$ decay, as introduced in Eq.~(\ref{S_3_K}).

The decay of  $\tau$  into three pseudoscalar mesons has been studied in literature~\cite{Decker:1993,Finkemeier:1995,Khun:1992}. %Our aim is to give an upper bound on the asymmetry $A_3$ to show that it can be safely neglected in experimental applications for Belle~II. 
For the present analysis, however, we aim only  at a rough estimate of the possible size of the effect.  We therefore adopt a simplified model for the process $\tau\rightarrow \nu K^0\overline{K}^0\pi$. Specifically, we treat the decay as proceeding through an intermediate $K^*$ resonance,  corresponding to a pseudo-three-body decay:
\begin{equation}
\tau^+\rightarrow \nu K^{*+}\overline{K}^0 \to 
\nu K^0\overline{K}^0\pi^+, 
\qquad 
\tau^+\rightarrow \nu K^0\overline{K}^{*+}\to 
\nu K^0\overline{K}^0\pi^+.
\end{equation}
The  $K^*$ resonance parameters are taken as~\cite{Decker:1993, Finkemeier:1995}:
\begin{equation}\label{K_star_res}
    m_{K^*} = 0.892\;\text{GeV},  \qquad \Gamma_{K^*} = 0.050\;\text{GeV}\;.
\end{equation}
Further details of the computation of the $K^0$ and $\overline{K}^0$ spectra in the $\tau$ rest frame are provided in Appendix~\ref{app:Kaon_spetrum}. 
We stress that this model is not intended to yield realistic predictions for the differential decay rate.  Instead, it represents an extreme scenario designed to maximize the difference between the $K^0$ and $\Kbar$  energy distributions. In this way, it serves to set an upper bound on the possible size of the experimental asymmetry $A_3$.

\begin{figure}
\centering
\includegraphics[width=0.8\columnwidth]{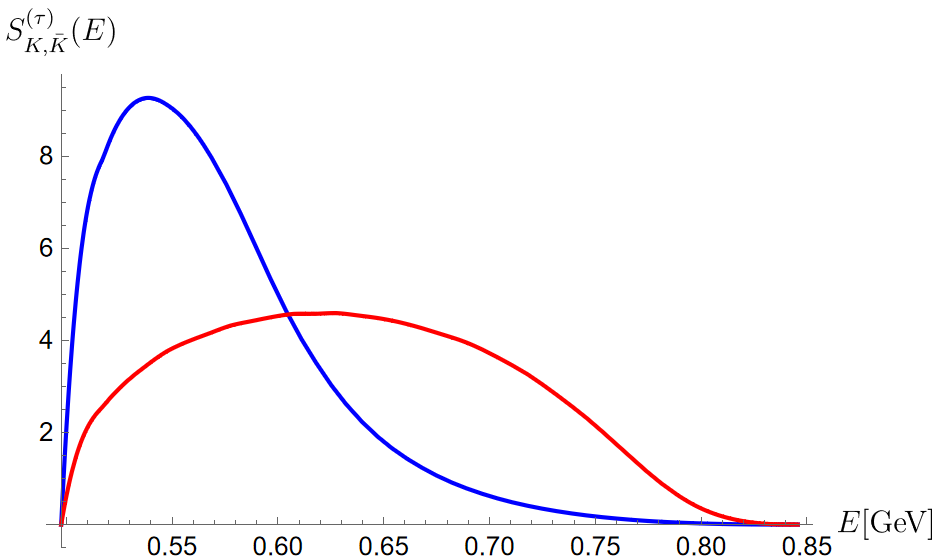}
\caption{Normalized energy spectra of the kaons in the decay $\tau^+\rightarrow \overline{\nu}\pi^+ K^0 \overline{K}^0$ in the $\tau$ rest frame: $S^{(\tau)}_K(E)$ for  $K^0$ (red line) and $S^{(\tau)}_{\overline{K}}(E)$ for  $\overline{K}^0$ (blue line). A simplified interaction model with an intermediate $K^{*}$ resonance is employed~\cite{Decker:1993, Finkemeier:1995}. 
%In particular, this model is a limit case in which the spectrum of the two kaons is more different (leading to an upper bound value for the asymmetry $A_3$).
}
\label{fig:spectrum}
\end{figure} 

We use the $\tau$ rest-frame spectrum shown in Fig.~\ref{fig:spectrum} and boost to the laboratory frame to obtain the kaon energy distributions $S_{K}(E)$ and $S_{\overline{K}}(E)$, following the same strategy used in the computation of $A_1$ and $A_2$ asymmetries (see Section~\ref{subsec:BELLEII_A1_A2}). Incorporating the detector efficiency given in Eq.~(\ref{F_BELLE2}), we compute
\begin{align}
    f_3^K(t) = \sum_{i=1}^{10} \int_{\Delta E_i} dE\; S_{K}(E)F_i(\beta\gamma t)\;, \\
    f_3^{\overline{K}}(t) = \sum_{i=1}^{10} \int_{\Delta E_i} dE\; S_{\overline{K}}(E)F_i(\beta\gamma t)\;.
\end{align}
Using Eq.~(\ref{A3_Final}), we obtain an estimate for the experimentally measured asymmetry:
\begin{align}\label{A3_result}
A_{Exp,3} \lesssim 
9.2 \times 10^{-5} \approx 0.03 \times 2\text{Re}[\epsilon]\;.
\end{align}
Within the framework of our simplified spectrum model, this result should be regarded as a rough upper bound. We therefore conclude that the $A_3$ asymmetry represents a very small effect, below the expected sensitivity of the Belle~II experiment.

\subsection{The total asymmetry}

The experimentally observed asymmetry $A_{Exp,tot}$ at Belle~II-type experiment  receives contributions from three decays:
\begin{equation}
    A_{Exp,tot}=\frac{R_1 A_{Exp,1} + R_2 A_{Exp,2} +R_3 A_{Exp,3}}{R_1+R_2+R_3}\,,
\end{equation}
where $R_{1,2,3}$ denote the event fractions of the corresponding decay modes in the total data sample.  The decays $\tau^\pm \to K^\pm K_S\, \nu$ and $\tau^\pm \to \pi^\pm K_S K_L \nu$  act as backgrounds to the $CP$ asymmetry measurement in $\tau\rightarrow \nu K_S\pi$ decay. Using the results of Eq.~(\ref{A12_result}) and Eq.~(\ref{A3_result}), we find that the following approximations hold to within a few percent accuracy:
\begin{equation}
A_2=-A_1, \qquad A_3=0  .  
\end{equation}
These simplifications are well justified within the expected experimental precision of Belle~II and can therefore  be used as working assumptions in the analysis of the measured asymmetry.

%%%%%%%%%%%%%%%%%%
\section{Additional effects}
\label{sec:additional_effects}

In this section, we  discuss  additional effects that may induce artificial contributions to the measured asymmetry, potentially mimicking the genuine $CP$-violating signal.

%We have checked that they are very small, and can be neglectected for the current stage of precision. Yet, in the future they might become relevant.

\subsection{Matter-induced effects in kaon oscillations} 

%The fact that the detector is composed of matter introduces corrections to the predicted $CP$ asymmetry. \desy{predictied or measured?}

The presence of the detector material introduces corrections to the measured $CP$ asymmetry.
One possible contribution arises from the different experimental reconstruction efficiencies of particles and antiparticles, which stem from their different interaction probabilities with the detector material. This effect is well established in the literature~\cite{BELLE2_paper_instrumental_asymmetry, Belle2:thech_2022instrumental}, and leads to a modification of the measured $CP$ asymmetry at the level of $10^{-3}$. Within our formalism, this contribution is included in $A_{Exp,P}^{\rm det}$, see  Eq.~(\ref{A1_Final}) and Eq.~(\ref{A2_Final}). 

Another effect, specific to neutral kaons, arises from matter-induced modifications of kaon oscillations, as described in Section~\ref{subsec:kaon_decay_rates}. In the presence of matter,  the effective mass eigenstates differ  from those in vacuum, as shown in  Eq.~(\ref{cSL_matter}).
Consequently, the measured $CP$ asymmetry $A_{Exp,P}^K$  receives a correction $A_{Exp,P}^{K,\text{mat}}$ relative to the vacuum case, see Eq.~(\ref{A_matter}). This regeneration effect modifies the effective time evolution of kaons in the detector, thereby shifting the observed $CP$ asymmetry relative to the vacuum expectation. As shown in Eq.~(\ref{Delta_pi_pi_Full}), this correction depends on the material-dependent kaon regeneration parameter $r$, defined in Eq.~(\ref{r_def}), which can be expressed as
\begin{equation}\label{r_DeltaA}
    r = i\pi n L\frac{\mathcal{A}-\overline{\mathcal{A}}}{p_K}\times\frac{1}{1/2-i\Delta m \tau_S}\;.
\end{equation}
Here $p_K$ is the momentum of the neutral kaon, and $L=\gamma\beta\tau_S$ is the average decay length of a $K_S$ in the laboratory frame. %The expression of $A_{Exp,P}^{K,\text{mat}}$ is found from the terms proportional to $r$ in Eq.~(\ref{Delta_pi_pi_Full}). 
In general, $r$ depends on the kaon energy in the laboratory frame, $r=r(E)$. Consequently,  the computation of $A_{Exp,P}^{K,\text{mat}}$ requires  averaging over the kaon energy in the laboratory frame,  as encoded in the efficiency function (see Eq.~(\ref{f_P})),
%As a consequence, to express the correction to the asymmetry, we need to explicitly include the average over the energy of the kaon in the laboratory frame that defines the efficiency function, Eq.~(\ref{f_P}), - I tried my best to polish this sentence, please check
\begin{equation}\label{A1_matter_analytical}
    A_{Exp,1}^{K,\text{mat}} = \frac{\int dE\;S_{1,K}(E)\int_0^{\infty}dt\;F(\beta\gamma t,E)\;\Delta_{\pi\pi}^{\text{mat}}(t,E)}{\int_0^{\infty}dt\; f^K_1(t)e^{-\Gamma_s t}}
\end{equation}
where we define:
\begin{equation}\label{Delta_pi_pi_matter}
    \Delta_{\pi\pi}^{\text{mat}}(t,E) = -2\text{Re}[r(E)]\bigg[ e^{-\Gamma_S t} - e^{-\Gamma t}\left( 
\cos(\Delta m t) + \frac{\text{Im}[r(E)]}{\text{Re}[r(E)]}\sin(\Delta m t) \right) \bigg]\;.
\end{equation}
Eq.~(\ref{A1_matter_analytical}) and Eq.~(\ref{Delta_pi_pi_matter}) are valid under the assumption that the entire decay length of the neutral kaon lies within  a material characterized with a single regeneration parameter $r$.  Eq.~(\ref{A1_matter_analytical}) can thus be viewed as a useful approximation, with $r$ computed as an average over different detector components. A fully realistic treatment requires solving the kaon equations of motion in the layered detector environment, see, e.g.,~\cite{Fetscher1996}.

As shown in  Eq.~(\ref{r_DeltaA}), the parameter $r$ depends on the kaons regeneration amplitudes $\Delta \mathcal{A}=\mathcal{A}-\overline{\mathcal{A}}$. Experimental measurements of $\Delta \mathcal{A}$ have been performed for different materials in the momentum range $p_K\in[20,140]\text{GeV}$~\cite{Gsponer:1979}, leading to the empirical relation:
\begin{equation}
\label{Df_empirical}
    \left| \frac{\Delta \mathcal{A}}{p_K} \right| = \frac{2.23\cdot A^{0.758}}{(p_K[\text{GeV}])^{0.614}}\;, 
\end{equation}
where $A$ denotes the nuclear mass number of the material. The corresponding phase $\text{arg}(\Delta\mathcal{A})$ was measured for Carbon in the range $p_K\in[35,125]\,\text{GeV}$~\cite{Roehrig}, and found to be consistent with the Regge theory prediction:
\begin{equation}\label{arg_Df}
    \text{arg}(\Delta \mathcal{A})\approx -\frac{\pi}{2}(2-0.614) = -124.7^{\;\circ}.
\end{equation}

These measurements, however, are obtained at kaon momenta much higher than those relevant for Belle~II, where $E\sim\mathcal{O}(1\text{GeV})$. At low energies, theoretical estimates of regeneration amplitudes have been presented in Ref.~\cite{Uchiyama:1994,Ko:2011}, and measurements exist for Carbon in the range $p_K\in[250,750]\,\text{MeV}$~\cite{ANGELOPOULOS1997422}. These results indicate a sizable theoretical uncertainty in the determination of $\text{arg}(\Delta \mathcal{A})$, with deviations from the Regge expectation in Eq.(\ref{arg_Df}) of an average of $35^{\circ}$  and up to $45^{\circ}$. Such deviations directly affect the determination of $r$, and consequently of $A_{Exp,1}^{K,\text{mat}}$, potentially even changing the sign of the matter-induced correction for a given $|r|$. 
%\desy{I am not sure what to conclude from this paragraph. Is this something we should worry about in this paper? Is this under control? Shall we exchange the next 2 paragraphs to make it more clear?}\gp{I think it's very important to point out that the matter correction can be $\sim100\%$ of the CP asymmetry (also is the BaBar anomaly reliable?). Figure \ref{fig:A1_matter} shows a theory estimate based on available measurements of $r$ at low energies, and we need it to motivate Eq. (\ref{eq:A_mat_BELLEII}), so I am not sure about exchanging the two paragraphs.}

\begin{figure}
\centering
\includegraphics[width=0.7\columnwidth]{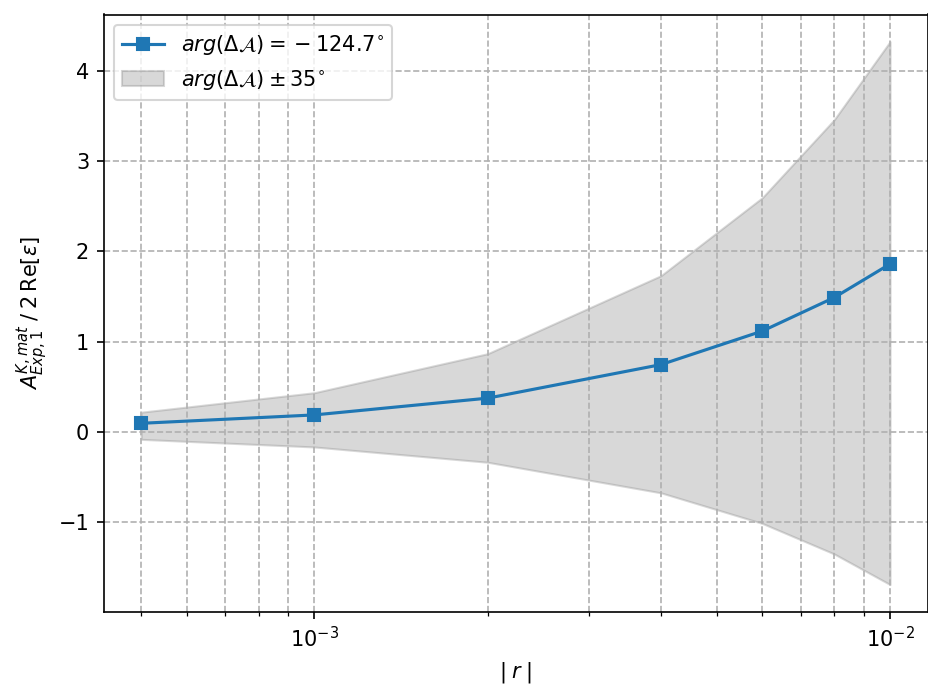}
%\caption{$A_{Exp,1}^{K,\text{mat}}$, Eq.~(\ref{A1_matter_analytical}), computed for different values of $|r|$, assuming $r$ to be constant (not dependent on the energy), and using the the experimental efficiency function of a Belle~II like experiment obtained in Section~\ref{sec:BELLE2}. The solid line corresponds to the case predicted by Regge Theory, Eq.~(\ref{arg_Df}), and the shaded region corresponds to a theoretical uncertainty of $35^{\;\circ}$ on $\text{arg}(\Delta\mathcal{A})$.} 
\caption{The correction $A_{Exp,1}^{K,\text{mat}}$, Eq.~(\ref{A1_matter_analytical}), 
%to the experimentally measured $CP$ asymmetry arising from matter-induced kaon regeneration effects, divided by the vacuum $CP$ asymmetry $\approx 2\text{Re}[\epsilon]$, 
plotted as a function of the magnitude of the kaon regeneration parameter $|r|$. The calculation assumes $|r|$ to be energy independent and employs the experimental efficiency function of a Belle~II obtained in Section \ref{sec:BELLE2}. The solid line corresponds to the phase predicted by Regge theory, see Eq.~(\ref{arg_Df}), while the shaded band reflects a theoretical uncertainty of $35^{\;\circ}$ on $\text{arg}(\Delta\mathcal{A})$.}
\label{fig:A1_matter}
\end{figure} 

In Fig.~\ref{fig:A1_matter}, we present a theoretical estimate of $A_{Exp,1}^{K,\text{mat}}$,  computed as a function of $|r|$, for different values of the theoretically uncertain phase $\text{arg}(\Delta\mathcal{A})$. The calculation employs the Belle~II efficiency function, as described in Section~\ref{sec:BELLE2}, and assumes, for simplicity, that $r$ is independent of the kaon energy.

In Ref.~\cite{Bjoern:2019qgw}, a theoretical estimate of $|r|$ was obtained for different types of detectors by averaging over their various components. For Belle~II, this gives $|r_{\text{Belle~II}}|\sim 10^{-3}$, which corresponds to a correction to the $CP$ asymmetry of
\begin{equation}\label{eq:A_mat_BELLEII}
    \left|A_{Exp,1}^{K,\text{mat}}\right|_{\text{Belle~II}}\sim 0.2 \times (2\text{Re}[\epsilon])\;.
\end{equation}
The uncertainty on this estimate is, however, very large and difficult to quantify reliably.  

Because in realistic detectors $r$ is of the same order as $\epsilon$, a precise determination of $r$ and its energy dependence is essential for accurate theoretical predictions of the measured $CP$ asymmetry. Moreover, since the effect is energy dependent, it may induce different apparent asymmetries for different decay modes.

\subsection{Forward-backward asymmetry} 
 
Another contribution to $A_{Exp,P}^{\rm det}$ arises in  asymmetric machines, such as the SuperKEKB $e^+e^-$ collider, where Belle~II operates. We discuss this effect for $A_1$. The generalization to other processes is straightforward.

The efficiency function $f^{\alpha}_P(t)$ depends on the energy spectrum of the kaon in the laboratory frame, see Eq.~(\ref{eq:genf}).
%In the case of the $A_1$ asymmetry we look at the process $\tau\rightarrow\pi K_S \nu$, and, as shown in Eq.(\ref{f_P}), the reconstruction efficiency is parametrized by the kaon spectrum $S_P(E)$ in the laboratory frame. 
%It is important to note that efficiency function $f_P$ not only depends on the decay process, but also whether the particle emitted is $K^0$ or $\overline{K}^0$. 
In general, the laboratory-frame energy distribution of a $K^0$ produced by a $\tau^+$, $S_{1,K}^+(E)$,  differs from that of a $\overline{K}^0$ produced by a $\tau^-$, $S_{1,\overline{K}}^-(E)$. This difference arises because the $\tau$ leptons are produced via  $e^+e^-\rightarrow \tau^+\tau^-$ processes, and the forward-backward (FB) asymmetry in the 2-to-2 scattering~\cite{Ferber:2016rka} leads to different angular distributions for the $\tau^+$ and $\tau^-$ in the center of mass frame, and a different energy spectra in the laboratory frame. Consequently, the resulting efficiency functions differ, $\Delta f_1^K \neq 0$, see Eqs.(\ref{f_sum}) and (\ref{eq:genf}). %This implies a different energy spectrum for the $K^0$ and the $\overline{K}^0$ in the laboratory frame, and, as a consequence, using Eq.~(\ref{eq:genf}), it implies that $\Delta f_1^K \neq 0$, see Eq.~(\ref{f_sum}). 
This effect gives a small correction to the measured asymmetry, see Eq.~(\ref{A1_Final}).
To estimate this effect for Belle~II, we consider the $\tau^+\tau^-$ production differential cross section~\cite{Ferber:2016rka},
\begin{equation}
    \frac{d\sigma_{ee\rightarrow\tau\tau}}{d\cos{\theta}}\propto F_1(1+\cos^2\theta) + 2F_2\cos\theta , 
\end{equation}
where $\theta$ is the scattering angle in the laboratory frame. The FB asymmetry is given by $A_{FB}=3F_2/4F_1$. The leading-order asymmetry can be computed from a one-loop calculation of the 2-to-2  scattering, which is beyond the scope of this work. Here, we assume $A_{FB}=1\%$~\cite{Ferber:2016rka, SOBIE:1999}, and use the Belle~II-like experimental efficiency and kaon spectrum from Section~\ref{sec:BELLE2}. Using Eq.~(\ref{A1_Final}), the resulting contribution of the FB asymmetry to the experimental asymmetry is
%Using Eq.~(\ref{A1_Final}) to compute the theoretical estimate of the experimental asymmetry $A_{Exp,1}^{\text{det}}$, we find that the effect of the FB asymmetry is
\begin{equation}
    A_{Exp,1}^{\text{det,\;FB}}\sim 10^{-3}\times \left(2\text{Re}[\epsilon]\right)\;,
\end{equation}
which is very small compared to the sensitivity of current experiments.

\section{Conclusion}

In the SM, $CP$ violation in $\tau\rightarrow\pi K_S \nu$ decays arises from neutral kaon oscillations in the final state. This $CP$  violation has been studied  by the BaBar and Belle collaborations, with the BaBar measurement showing a slight tension with the SM prediction. The Belle~II collaboration aims to provide a more precise measurement of this $CP$ asymmetry. In this work, we present a method to predict the experimentally measured $CP$ asymmetry in $\tau$ decays into final states containing neutral kaons.% In this work, we showed how to make the theoretical prediction for the experimental measured $CP$  asymmetry in the decay of the $\tau$ into final states containing neutral kaons. 

We note that the theoretical framework presented here is more general. It can be applied to any decay into a final state containing neutral kaons, such as charm decays $D^+\rightarrow K_S \pi^+$.

%We remark that the theory discussed in this work to provide a theoretical prediction for the measured $CP$  asymmetries in $\tau$ decays into neutral kaons is more general, and it applies for any decay into a final state containing neutral kaons, such as charm decays $D^+\rightarrow K_s \pi^+$. 

Below, we summarize the main results of this work:

\begin{enumerate}
\item
Theoretical predictions for the $CP$ asymmetries require knowledge of the experimental reconstruction efficiency, $f_{P}^{\alpha}(t)$, as a function of the kaon decay time in the kaon rest frame. In section~\ref{sec:efficiency_function}, we show that the efficiency function has two contributions, see Eq.~(\ref{eq:genf}): $(i)$ the kaon energy distribution in the laboratory frame, which depends on the specific production process $P$, and $(ii)$ a universal reconstruction efficiency function that depends on the features of the detector and the experiment. 
%As a consequence, the measured $CP$  asymmetry $A_{Exp, P}$ depends on both the specific process and the detector and the analysis details. 
\item
In section~\ref{sec:A3}, we show that because the reconstruction efficiency depends on the kaons energy spectrum, the decay channel into two neutral kaons, $\tau\rightarrow \nu K_L K_S \pi$, exhibits $CP$ violation, see Eq.~(\ref{A3_Final}). This arises from the fact that the two neutral kaons are produced with different energy distribution in  $\tau$ decays. 
\item
In section~\ref{sec:BELLE2}, we provide a theoretical prediction for the measured $CP$  asymmetries for the processes $A_1$, $A_2$ and $A_3$ in a Belle~II-like experiment. We find that, within a few percent, it is justified to approximate $A_2  \approx -A_1$ and $A_3  \approx 0$.

%We performed a simulation of the Belle~II detector and derived an estimate of the efficiency functions parameterized by the decay time and kaon for all the different processes. We found the numerical estimates for the asymmetries $A_{Exp,1}\approx3.00\times 10^{-3}$ and $A_{Exp,2}\approx -3.07\times 10^{-3}$, and we estimated a theoretical upper bound on the experimental value of the asymmetry $A_{Exp,3} \lesssim 9.2\cdot 10^{-5}\approx 0.03\times 2\text{Re}[\epsilon]$.
\item
In section~\ref{sec:additional_effects}, we find that matter effects on neutral kaons oscillations can produce a significant correction, up to order $\mathcal{O}(1)$, to the measured $CP$ asymmetry, see Eq.~(\ref{A1_matter_analytical}) and Fig.~\ref{fig:A1_matter}. 
%Such correction depends on the kaons regeneration parameter $r$, Eq.~(\ref{r_def}), which depends on the properties of the material and the energy of the neutral kaon. 
%Using the limited information on kaons regeneration available in the literature, we estimated that the correction to the $CP$  asymmetry due to kaons oscillation in matter can be very important, 
%up to order $\mathcal{O}(1)$ in realistic experiments. However, there is a significant theoretical uncertainty on the parameter $r$ at low energies, and 
Further studies and measurements are required to provide a precise theoretical prediction of this correction.
\end{enumerate}

We conclude with the hope that our results will be useful for the upcoming experimental analyses of $CP$‑violating effects in final states with neutral kaons.

\section*{Acknowledgments}

YG is supported in part by the NSF grant PHY-2309456.
PL, AM and AR are supported by the German Ministry of Education and Research (BMBF), %the Deutsche Forschungsgemeinschaft (DFG),  
and the Helmholtz Association (HGF).

\appendix
\section{Kaons energy spectrum in the $\tau$ frame} \label{app:Kaon_spetrum}
In this appendix we discuss the simplified theoretical models used to compute $S_{\alpha,P}^{(\tau)}$ , the energy spectrum of the kaon emitted in the $\tau$ frame for the processes considered.

\textbf{$\tau$ decay into two mesons:} We consider the decay process $\tau^+\rightarrow \overline{\nu} M^+ K_S$, where the meson $M^+=\pi^+,\;K^+$ for $A_1$ and $A_2$, respectively. To describe the decay, we use a simplified model of a point-like interaction between real scalar fields,
\begin{equation}
\mathcal{H}_{int}=g\;\overline{\tau}(x)\nu(x) M^+(x) K(x),
\end{equation}
where $g$ is a coupling constant, and $K(x)=K^0(x)$ for process $A_1$, and $K(x)=\overline{K}^0(x)$ for process $A_2$. In this simplified model, the scattering matrix element $\mathcal{M}=\bra{M^+K^0\overline{\nu}}\mathcal{H}_{int}\ket{\tau^+}$ is constant, $\mathcal{M}=g$. The normalized kaon energy spectrum  is obtained from the normalized differential rate as a function of the kaon energy. Since the scattering matrix element is constant, the kaon's energy distribution essentially depends only on the phase space factor in the differential rate. %$\frac{d\Gamma}{dE_K}$ for the different processes.
\begin{comment}
The decay rate is given by
\begin{equation}
    d\Gamma = \frac{1}{2m_{\tau}}|\mathcal{M}|^2(2\pi
)^4\delta^{(4)}(p_{\tau}-p_{\nu}-p_{M}-p_K)\frac{d^3p_{\nu}}{(2\pi)^3E_{\nu}}\frac{d^3p_{M}}{(2\pi)^3E_{M}}\frac{d^3p_{K}}{(2\pi)^3E_{K}}
\end{equation}
\end{comment}

\textbf{$\tau$ decay into three mesons:} We consider the decay of  $\tau$ into three pseudo-scalar mesons $\tau^+\rightarrow \overline{\nu}\pi^+K^0\overline{K}^0$. To compute the kaons spectrum, we use a simplified model of a four-body decay interaction between real scalar fields, including the propagator of an intermediate $K^*$ resonance~\cite{Decker:1993, Finkemeier:1995}. The interaction is given by
\begin{equation}
\mathcal{H}_{int}=g_1\;\overline{\tau}(x)\nu(x) \Kbar(x) K^{*+}(x) + g_2\;K^{*+}(x)\pi^+(x)K^0(x)
\end{equation}
The differential decay rate is  obtained by treating the $K^*$ as an intermediate resonance:
\begin{equation}
{\cal M} = \frac{g_1g_2}{q^2 - m_{K^*}^2 + im_{K^*}\Gamma_{K^*}}\,.
\end{equation}
\begin{comment}
that leads to
\begin{equation}
d\Gamma = \frac{1}{2m_{\tau}}\frac{g^2}{|q^2 - m_{K^*}^2 + im_{K^*}\Gamma_{K^*}|^2}\frac{1}{(2\pi)^5}\delta^{(4)}(p_{\tau}-p_{\nu}-p_{K}-p_{\overline{K}}-p_{\pi})\frac{d^3p_{\nu}}{E_{\nu}}\frac{d^3p_{\overline{K}}}{E_{\overline{K}}}\frac{d^3p_{K}}{E_{K}}\frac{d^3p_{\pi}}{E_{\pi}}
\end{equation}
where, we defined $q=p_{\tau}-p_{\overline{K}}-p_{\nu}$, and $m_{K^*},\Gamma_{K^*}$ are defined in Eq. (\ref{K_star_res}). 
\end{comment}
The spectra for $K^0$ and $\overline{K}^0$ are defined as the normalized differential rates $\frac{d\Gamma}{dE_K}$ and $\frac{d\Gamma}{dE_{\overline{K}}}$.

%%%%%%%%%%%%%%%%%%%%%%%%%%%%%%%%%%%%%%%%%%%%%%%%%%%%%%%%%%
% \appendix
\nocite{*} 

%\bibliographystyle{unsrtnat}
%\bibliographystyle{utphys28mod}

%\bibliographystyle{unsrt}
%\bibliography{biblio}

%This for titles
%\bibliographystyle{unsrtnat}
%\bibliography{biblio}

%\nocite{*} %gives error of multiple inclusions 
\bibliographystyle{apsrev4-2}    % APS style
\bibliography{biblio}  

%\bibliographystyle{abbrv}
%\bibliography{biblio}

%\nocite{*}            % list all entries in biblio.bib
%\printbibliography
%%%%%%%%%%%%%%%%%%%%%%%%%%%%%%%%%%%%%%%%%%%%%%%%%%%%%%%%%%%

\end{document}